# Assessing the feasibility range of Solar-powered Planetesimals Redirection operations for Terraforming

https://drive.google.com/file/d/1kp2-FntYp4qx4eczv8tCuvY6W4kxWMSI/view?usp=sharing

Yegor A. Morozov*[ab], Mahdi Yoozbashizadeh[a], Ahmad Bani Younes[c], Saeid Janani[a], Mikhail Bukhtoyarov[d], Sergey V. Trifonov[e], Bryant K. Beeler[f], Ericson López[g]

* Expansist@gmail.com

[a] Mechanical & Aerospace Engineering Department, California State University Long Beach, 1250 Bellflower Blvd., Long Beach, California, 90840, USA

[b] Institute for Mathematical Sciences, Claremont Graduate University, 150 E. 10th Street, Claremont, California, 91711, USA

[c] Aerospace Engineering Department, San Diego State University, 5500 Campanile Drive, San Diego, California, 92182, USA

[d] Faculty of Liberal Arts and Sciences, I St, Budva, 85310, Montenegro

[e] Institute of Biophysics SB RAS, Akademgorodok 50 Corpus 50, Krasnoyarsk, Krasnoyarsk Krai, 660036, Russia

[f] Purdue University, 610 Purdue Mall, West Lafayette, Indiana, 47907, USA

[g] Escuela Politecnica Nacional, Quito Astronomical Observatory, Facultad de Ciencias, Departamento de Física, E11-253 Av. Ladrón de Guevara, Quito, 170143, Ecuador

## Highlights

- Nuclear fuel is scarce in Cosmos, must be saved for Interstellar Colonization missions where it is irreplaceable
- Thus Solar-powered Planetesimals Redirection for Terraforming technologies must be developed to scale up
- Precisely beaming solar power for Ion Thrusters using Planetesimal material as propellant is possible in entire Kuiper Belt with present technologies, to Oort cloud with 1-3+ order of magnitude precision increase

## Abstract

The work aims to assess principal physical and technical constraints of the key technological concept for achieving maximal carrying capacity and expansive potential of Planetary Systems. Chemical compositions of most potential Biosphere Substrates ($0.3*g_{Earth}<g_{BS}<3*g_{Earth}$ terrestrial (exo)planets and gas giant big (exo)moons) are in most cases very far from optimal for full-scale habitability thus productivity potential realization. This means, to cover all surface of a Biosphere Substrate with a dense photosynthesising layer during most of the remaining Planetary System lifecycle, for achieving maximal carrying capacity thus workforce and productivity enabling further expansion of the biosphere to new substrates - importing significant amounts of the limiting chemical elements by redirection of planetesimals material is ubiquitously inevitable. Although the price of such operations is high - the value of maximal population and productivity of a Biosphere Substrate during the remaining Planetary System lifecycle is orders of magnitude higher. According to orbital mechanics laws, the farther a planetesimal is orbiting from a star, the more time but less energy is required for its delivery. Also the more distant from a star, the more rich in volatiles planetesimals are. The most technically detailed and feasible scheme to date of *Planetesimals Redirection* was proposed in 1993 by Dr. Zubrin based on thermonuclear engines like those tested in the N.E.R.V.A. project. Although nuclear fuel is extremely scarce and might be the principal limiting factor for final stages of interstellar colonization missions. We investigate principal limitations and possibilities of using concentrated beamed solar energy to power *Planetesimals Redirection* operations for *Terraforming*, as it can increase the possible scale of such operations to orders of magnitude. Such a mission might consist of a “Planetesimal Tug” Spacecraft System (PTSS) to redirect a planetesimal, and a solar Power Harvesting & Beaming System (PHBS) concentrator, tracking tug spacecraft with a beam to power it. When the PTSS reaches a targeted planetesimal and mounts on it, covering the planetesimal with photoreceivers also decreases its evaporation from solar wind and radiation. The propulsion system might evolve from steam rockets to ion thrusters up to relativistic particle accelerators, increasing its technical complexity and efficiency, thus consuming less fraction of a planetesimal as propellant. The PHBS concentrator might be based on a momentum damper (Mercury), quite a limited surface. Or, on orbital Fresnel lenses with optic fiber flexible collimators, which might operate as arrays, allowing orders of magnitude increased surface area to collect starlight power. PHBS mostly passive orbit maintenance and tracking the PTSS during the entire mission are rather complex astrodynamical tasks we propose and estimate solutions for. Such kinds of operations principal limitations include pointing and tracking accuracy, with respect to increasingly delayed feedback, beam divergence, and more, increasing with interplanetary distance range. The feasibility of some proposed solutions can be tested with near-term missions. While deeper fundamental understanding and development of technical solutions to increase the feasible distances of solar powered *Planetesimals Redirection* operations can have profound effect on *Terraforming* research, making the field much more feasible and attractive, accelerating the works for Cosmic Expansion of our future generations.

## 1. Introduction

The goal of the study is to investigate principal technical possibilities of increasing long-term planetary habitability, carrying capacity thus workforce and industrial productivity for further expansion of our posterity in generations and our Biosphere to new planetary bodies.

Terraforming a planet is basically maximizing its carrying capacity for our posterity in generations during the remaining Planetary System lifecycle, so it can provide habitable environment, breathable atmosphere, water, food and everything else needed for reproduction of maximal number of people thus workforce [Morozov et al 2021a; Zubrin et al 1993; Birch 1991, 1992; Pazar 2017; Hossain et al 2015; Brandon 2018; Visysphere Mars 2005; Oberg 1981]. A Biosphere Substrate is any celestial body suitable for Terraforming & full-scale long-term inhabitation by our posterity in generations [Morozov et al, 2018a]. Biosphere Substrates are mostly Terrestrial Planets & smaller edge of gaseous or icy planets, which includes Subterrans (Mars size), Terrans (Earth size), Superterrans (Super-Earths & Mini-Neptunes), & in some cases Neptunians (lower range of Neptune size), but may also include some biggest moons of Jovian planets, whose size & gravity is comparable to Terrestrial Planets - Biosphere Substrates are basically most celestial objects within surface gravity range of $0.3*g_{Earth} < g_{BS} < 3*g_{Earth}$ orbiting within a technically extendable Circumstellar Habitable Zone at any time during the Planetary System lifecycle [Morozov et al, 2018a; Deeg et al, 2018; PHL, 2018; Fu et al, 2025].

The amount of biological cells thus organisms that can be assembled from any substrate, in our case from a planet (Biosphere Substrate), is always determined by the amount of the limiting chemical element quantity [Morozov et al, 2018a]. The element that is relatively most lacking in biological stoichiometric proportion always determines the upper limit of numbers of organisms that can assemble from a Biosphere Substrate & live on it, no matter how much of each other chemical elements are available - it is always about the limiting one. Thus, the biggest, most difficult & most important work in the whole Terraforming process, that determines the population numbers for the rest of a Planetary System lifecycle - is to deliver sufficient amounts of all the limiting chemical elements to all the Biosphere Substrates, as once you deliver a lot of one limiting chemical element - some other element becomes the limiting. In Earth's agriculture, the limiting chemical elements are mainly N & P, but on other planets these might be different chemical elements. Venus has lost almost all of its hydrogen, and water is the basic juice of life - so H will be the main limiting chemical element on Venus. Mars has about 2 orders of magnitude less dense atmosphere than Earth, while even if we melt everything on Mars - we can maximum increase Martian atmospheric pressure from present less than 1% to up to about 7% [NASA, 2018] of Earth's atmospheric pressure. In order to Terraform Mars, we need to import planetary amounts of H, N, C, P, & all other biogenic elements - basically we need to import the atmosphere & hydrosphere. Thus, Terraforming Venus & Mars is not feasible with chemical elements only present on them now, and requires importing planetary amounts of the limiting chemical elements in each case. Similar situation is with most ExoPlanetary Systems - present chemical composition of most Biosphere Substrates in observable part of Cosmos is usually far from optimal for maximal productivity of life, both in this & most other Planetary Systems [Deeg et al, 2018; Tinetti et al, 2018; Wordsworth, 2022; Zi-xin & Jiang-hui, 2024; Fu et al, 2025], thus requires work of much more numerous population than humankind currently has, during many generations, to cultivate, configure, Terraform, to make them habitable for our proliferating posterity in generations.

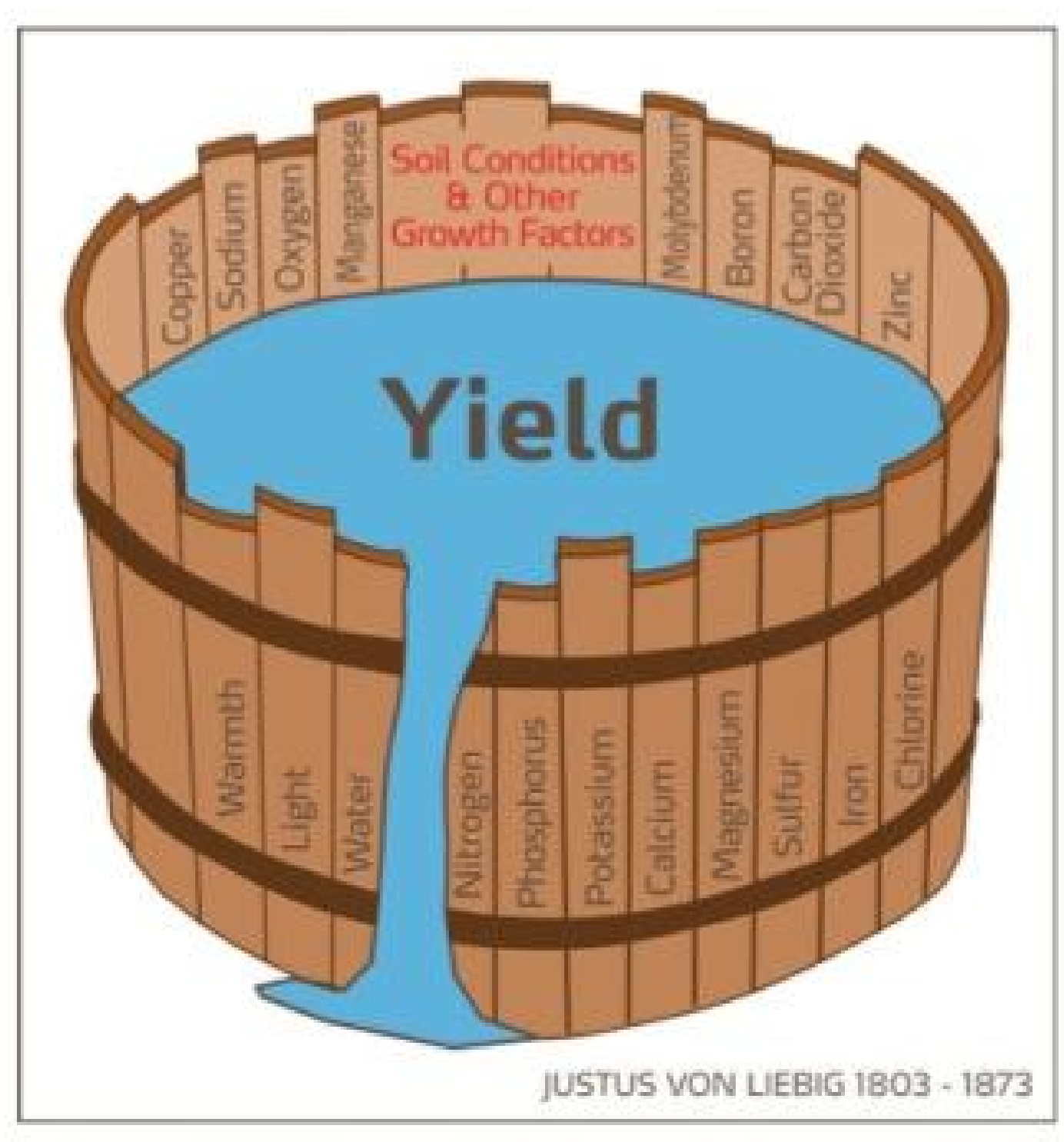

Fig. 1. Liebig's barrel, graphical visualization of the "Limiting chemical element" concept [https://earthwiseagriculture.net/grower-s-toolbox/law-of-minimums/]

Energy is produced either by fusion of light elements like Hydrogen, or by fission of heavy elements like Uranium. Middle mass chemical elements, from which most rocky planets are composed of - are mostly stable, they don't make nuclear reactions unless you pump into them much more energy than they release. Even if we stop thinking for a moment about those incredibly high temperatures & pressures available only inside stars, without which nuclear reactions of non-negligible scale don't occur, & other technical difficulties making scaling down a fusion reactor to a size less than a natural star in economically efficient way almost impossible

- even theoretically physically there is no way to make nuclear reaction producing planetary scale amounts of H from Fe, Si, C, & other middle mass elements energetically cheaper than it takes to deliver them already existing by transporting volatile rich remote Planetesimals from outer orbits where they are abundant, to inner planets where they are needed. As there is no economic way of making H, N, & other life ingredients from heavier elements - the cheapest & most realistic way to provide maximal biological & industrial productivity over a remaining Planetary System lifecycle is to import the limiting chemical elements by Planetesimals Redirection operations, prior to any other Terraforming operations. Thus, importing the limiting chemical elements by Solar-powered Planetesimals Redirection for Terraforming - is ubiquitously the biggest & most important part of all the Terraforming processes, which determines planetary habitability to the most extent [Morozov et al, 2021a].

Fig. 2. Presently known Kuiper Belt objects orbits. Credit: IAU MPC.

Presently known Kuiper Belt Objects orbits, where volatile rich Planetesimals start, beginning at about 30 AU and farther from the Sun, are mapped on Fig. 2. Their colors are diverse, mostly from gray and white to red and brown, which indirectly indicate high diversity of their chemical compositions. Their densities range from less than 0.4 to more than 2.6 g/cm$^3$, suggesting that Planetesimals with more than half mass fraction of $H_2O$, $NH_3$, and other valuable volatile ices exist in significant amounts [Jewitt et al, 1998, 2001, 2004; Delsani et al, 2006; Bockelée-Morvan et al, 2004; Mumma et al, 2011; Vernazza et al, 2016; Morbidelli, 2006; Vilenius et al, 2014; Bernstein et al, 2004; Fernández, 2020] The least dense objects are thought to be largely composed of mostly valuable ices and have significant porosity [Fernández, 2020; Jewitt et al, 1998]. The densest objects are likely composed mostly of rocks & metals with a thin crust of ice [Brown et al, 2012]. The detailed spectroscopic observations allowing significant signal-to-noise ratio of astrochemical measurements were done only to several dozens of Trans Neptunian Objects (TNOs), yet we already observe such unexpected diversity in chemical compositions that scientists have difficulties explaining [Jewitt et al, 1998, 2001, 2004; Delsani et al, 2006]. Most abundant volatile chemical compounds in TNOs are usually water ices, followed by carbohydrates and ammonia ices [Brown et al, 2012; Vernazza et al, 2016], but exact average % are yet difficult to estimate. There is a trend of low densities for small objects and high densities for the largest objects. One possible explanation for this trend is that ice was lost from the surface layers when differentiated objects collided to form the largest objects [Brown et al, 2012]. This trend can suggest that smaller icy Planetesimals about units of kilometers in diameter might be relatively more rich in volatiles on average then bigger Planetesimals about dozens to hundreds kilometers in diameter, thus most Planetesimals preferred to be redirected for Terraforming might be about units of kilometers in diameter.

Yet, it is clear that we have observed only a small fraction of the Kuiper Belt Objects (KBOs) and other Trans Neptunian Objects (TNOs) to date, mostly the biggest of them [Bernstein et al, 2004]. The presently observed trend is that the number of Planetesimals beyond Neptune tend to have ratio of numbers of Planetesimals versus their corresponding diameters proportional to their diameter to the q = -4 ± 0.5 power [Bernstein et al, 2004] (Fig. 3), which suggests there are millions of smaller objects & total mass of smaller size TNO Planetesimals is much bigger than total mass of bigger size Planetesimals, that is why we are mostly focused on delivering a few km diameter and smaller size objects to Mars & Venus, as they tend to have higher % mass fraction of valuable of $H_2O$, $NH_3$, & other volatiles ices.

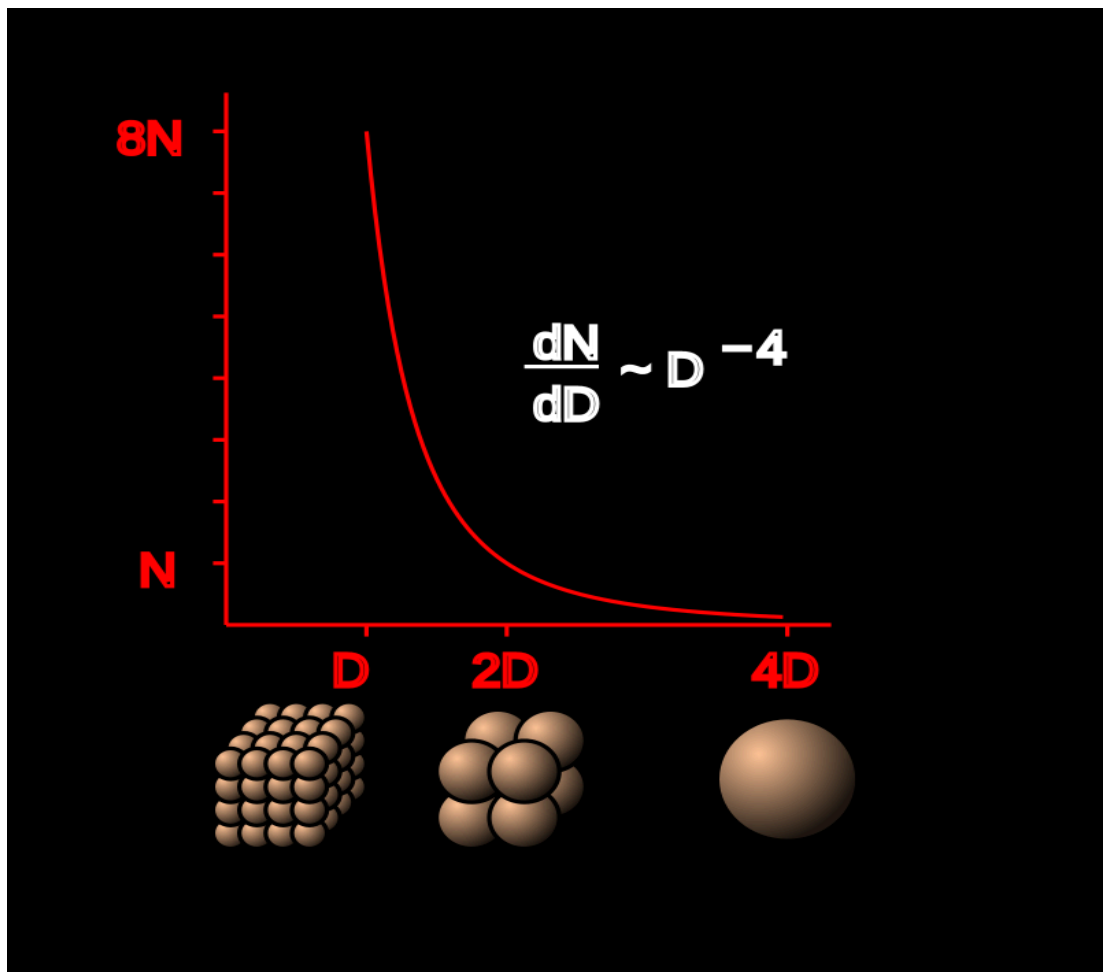

Fig. 3. Trend of observed TNOs numbers versus their sizes [Bernstein et al, 2004].

It is not easy to find the average composition of TNOs, as most papers are about a particular KBO composition; only several dozens TNOs have been chemically analyzed with spectroscopy in detail. Yet, some more detailed information is present regarding comets' chemical composition, because they have highly elliptical orbits and periodically approach closer distances to Earth, making more detailed chemical analysis easier for us [Bockelée-Morvan et al, 2004; Mumma et al, 2011; Vernazza et al, 2016].

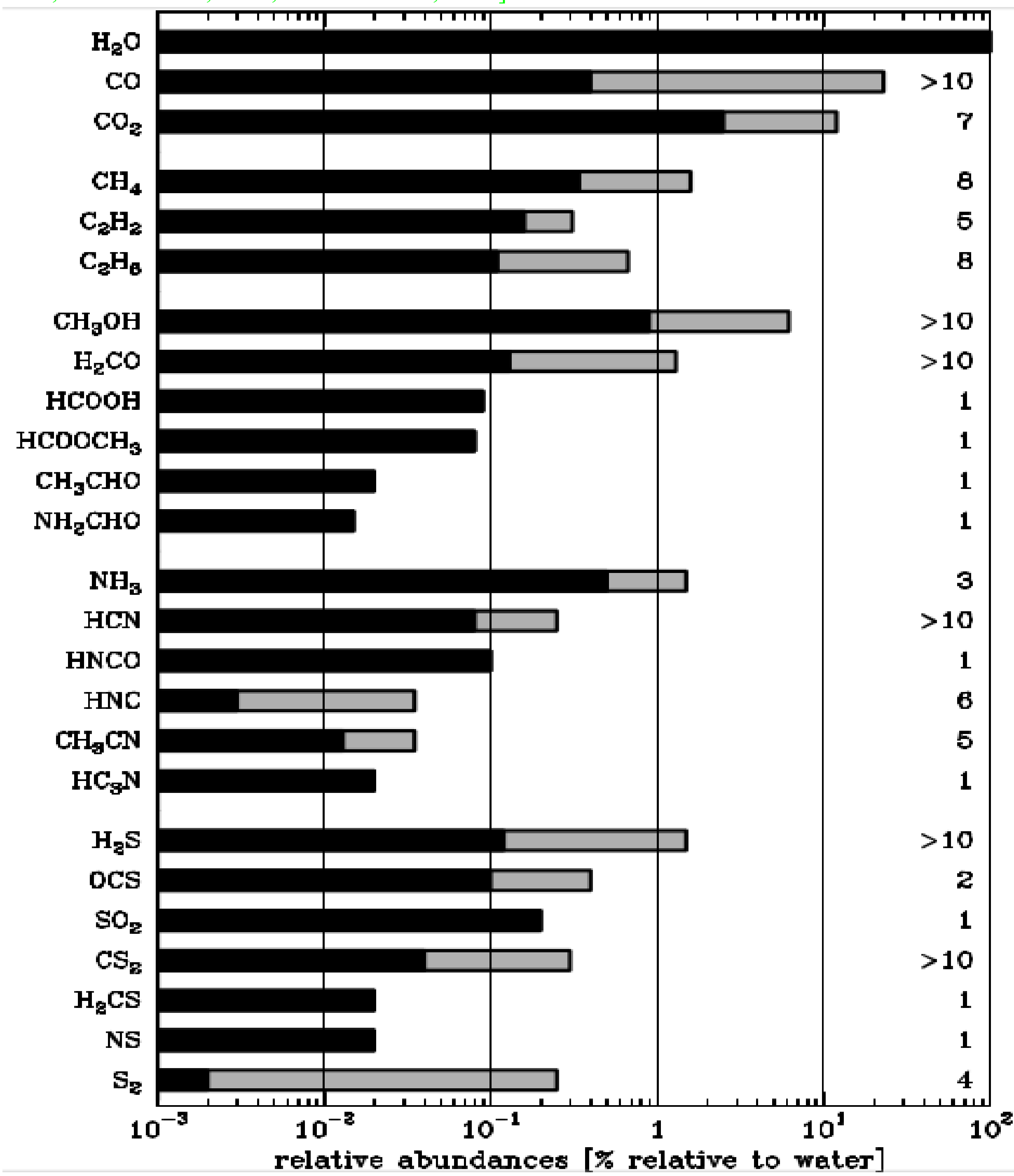

Fig. 4. Relative proportion of abundances of volatile compounds in comets analyzed yet [Mumma et al, 2011]

More than 20 volatile chemical compounds have been detected in comets by spectroscopic analysis [Bockelée-Morvan et al, 2004; Mumma et al, 2011; Vernazza et al, 2016] (see Fig. 4), but as we had detailed spectroscopic measurements of only several dozens out of probably millions existing and thousands detected TNOs – it is still rather difficult to estimate true average volatile chemical compounds proportion percentages due to yet small number of measurements, available data set is not statistically representative yet.

From yet known, very roughly, we can consider that the average proportion of volatile chemical compounds in Planetesimals beyond Neptune can be, if taken water $H_2O$ as 100% – then we would have about 10% of carbon oxides $CO_2$ + CO; about 3% of total various carbohydrates $C_xH_yO_z$; about 1% of ammonia $NH_3$; about 0.3% total sulfur S containing compounds [Bockelée-Morvan et al, 2004; Mumma et al, 2011; Vernazza et al, 2016].

Icy Asteroids, Comets, & other Planetesimals' materials on outer orbits of Planetary Systems [Deeg et al 2018; Marino, 2018; Pavlenko et al, 2021; Strom, 2024; Fu et al, 2025], beyond the orbit of Neptune in our case, are very numerous, abundant, very diverse in orbital parameters, sizes, masses, & chemical content [Morbidelli, 2006; Jewitt et al, 1998, 2001, 2004; Delsani et al, 2006]. Although each of them is relatively small - about units of kilometers in diameter or less on average - there are millions of them, their total mass is estimated to be more than mass of a terrestrial planet [Morbidelli, 2006; Bernstein et al, 2004; Fernández, 2020], which allows to assemble surface chemical composition optimal for life to develop a dense photosynthesising layer on all available surfaces of all Biosphere Substrates in a Planetary System [Jewitt et al, 1998, 2001, 2004; Delsani et al, 2006].

Thus, the main principal limiting factor for providing surface chemical composition optimal to achieve maximal biological and industrial productivity of a Planetary System during all its remaining life cycle time, is the amount of energy required for delivering planetary amounts of the limiting chemical elements by Planetesimals Redirection operations for Terraforming. Also Planetesimals Redirection operations can adjust planetary rotation, period = daylength & tilt for more even surface irradiation thus productivity & climate, by striking Planetesimals' tangential to each Biosphere Substrate desired Equator with high velocity, strengthen planetary magnetic fields by assembling satellites of significant size that create constant tidal forces enforcing current conducting fluid circulation flow inside molten planetary interiors & creating inhomogeneities, where heating Planets including interiors is just a useful side effect, & help other Terraforming tasks. This technology development can also make all planetary defense issues easy as the result of having PTSSs fleets constantly orbiting & observing, as another useful by-product.

Biological seeding is the easiest, cheapest, and most straightforward part of Terraforming that was already studied & discussed in detail in older publications [Fogg, 1998; Beech, 2009; Pazar, 2017], so we won't dedicate much volume for biological seeding questions in this publication. The general principle of biological seeding of a Terraformed Planet is introducing organisms gradually from simplest smallest to larger & more complex - first bacteria, unicellular photosynthesising like *chlorella vulgaris* and nitrogen-fixing bacteria to make initial organic material, oxygen & proto-soil; second grass & legume seeds planted once proto-soil created to create primary soil, third bushes fruit trees big productive plants are seeded to create breathable atmosphere & food, fourth animals including fish and human are introduced to newly Terraformed Biosphere Substrate, which was already well described before us [Fogg, 1998].

The most technically detailed description yet of Planetesimals Redirection operations for Terraforming was published by Dr. Zubrin in 1993 [Zubrin et al, 1993], based on the quartet of N.E.R.V.A. type 5000 MWt NTR (Nuclear Thermal Rocket) engines, with total power 4*5 GW = **20 GW**, calculating the mass, energy, & time requirements for a **2.6 km** diameter **$10^{13}$ kg** ammonia-rich asteroid delivery to Mars depending on its initial orbit. According to Kepler's all 3 laws of orbital mechanics [Curtis, 2014; Vallado, 2001], the more distant the orbit semi-major axis is from the most massive = central body (star) - the less is the orbital velocity & specific orbital energy of an object at that stable orbit, more time but less energy is required to deliver a Planetesimal to the destination Biosphere Substrate from more distant orbits, especially if we use all the Gravity Assist Maneuvers potential around all the intermediate Gas Giant Planets, to save propellant & time. Also, the more distant from a star - the more rich in volatiles the planetesimals are on that orbits, as well as more numerous, and their total mass corresponding to longer orbits thus bigger volumetric segments of space is generally higher. That is why increasing the technically possible range of Solar-powered Planetesimals Redirection operations for Terraforming is very important.

## 2. Materials and methods

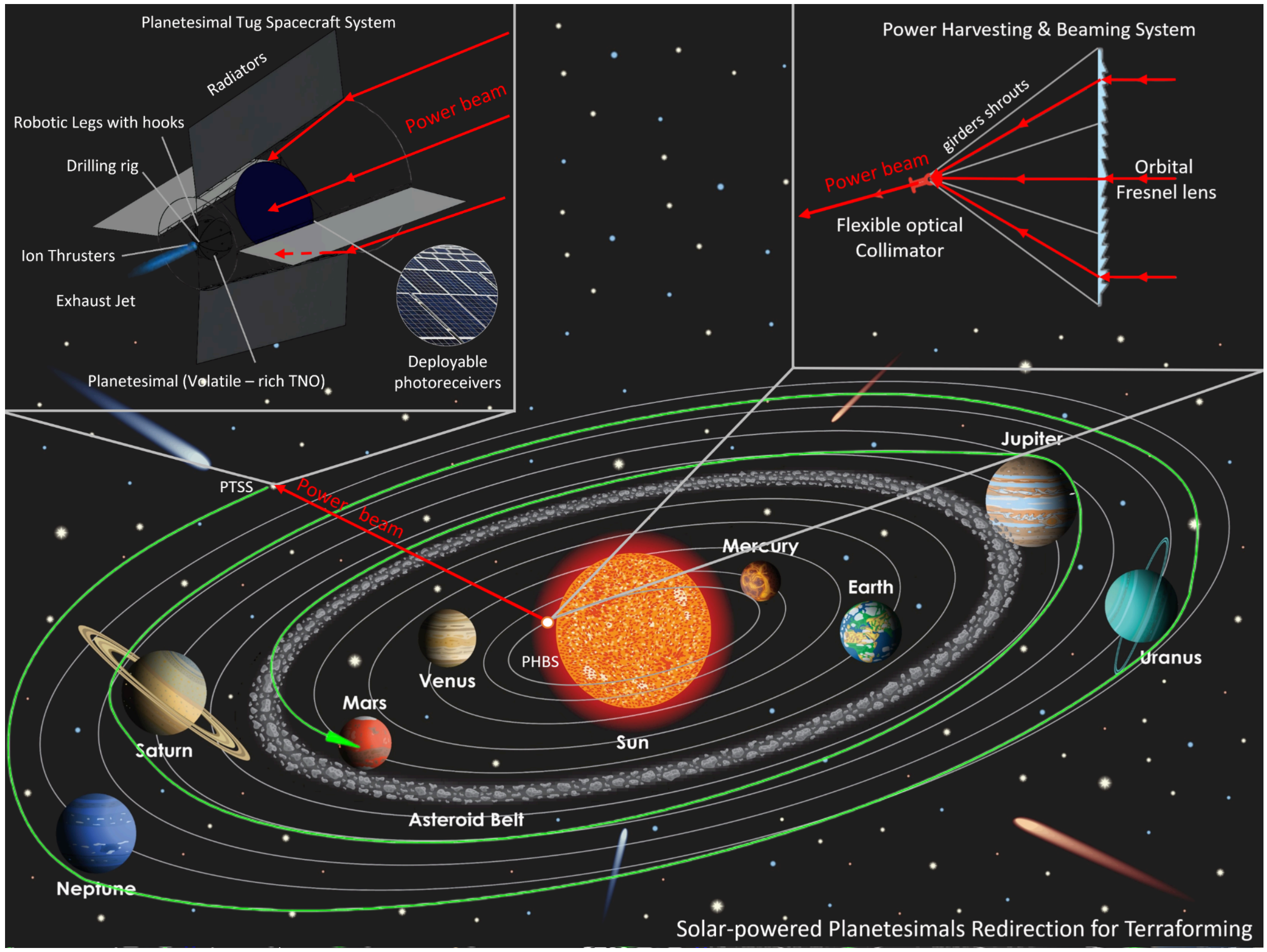

Fig. 5. Solar-powered Planetesimals Redirection for Terraforming concept overview

A Planetesimal Redirection mission architecture (see Fig. 5) might consist of:

1. A “Planetesimal Tug Spacecraft System” (PTSS), with solar-powered deep space electric propulsion system, landing springs for soft landing, robotic legs with hooks to mount rigidly on the planetesimal & control attitude, expendable drilling rig with telescopic segments to extract & feed propellant, and deployable photoreceivers covering the targeted planetesimal upon arrival & landing on it, providing electric power for propulsion and other systems.

2. A “Power Harvesting & Beaming System” (PHBS) based on ultra-thin ultra-lightweight lucid nano-foil concentric rings orbital Fresnel lens with 2-axis gimballed flexible optical fiber collimator and/or 2 axis gimbaled solar-pumped laser on the guiding end [Forward, 1984; Frisbee, 2004], tracking the PTSS 2-axis gimbaled photoreceivers photovoltaic array with concentrated light beam to power it continuously during all the entire mission.

The PHBS lens system circular heliocentric orbit radius is determined by 2 factors: the closer it is to the star the less surface area is required to collect the same power amount, decreasing the system mass & cost in inverse square proportion from the orbit semi-major axis - but proximity to a star is limited by the temperature at which PHBS can maintain the mechanical properties and control systems reliable operation. The power harvesting sunlight concentrators might be based on a momentum dampher (Mercury), which is quite a limited surface. Or, on orbital Fresnel lenses with 2-axis gimballed optic fiber flexible collimators and/or solar-pumped lasers, which might operate as arrays, which can allow orders of magnitude increased surface area to collect solar power. The PTSS propulsion system might evolve from nuclear steam rockets to ion thrusters up to relativistic particle accelerators, increasing technical complexity & efficiency - consuming less fraction of planetesimals as propellant. Though Ion Thrusters are most preferable because they can use almost any gas as propellant, use solar power as energy source, and are very propellant efficient, highest $I_{sp}$ interplanetary jet propulsion system known today - Ion Thrusters exhaust velocity is limited only by power available [Goebel et al, 2008; Yamamoto et al, 2018].

A most common Solar-powered Planetesimal Redirection for Terraforming mission concept might be following:

The PTSSs’ & PHBSs’ parts are launched from Earth (or any other start Biosphere Substrate) Equator surface to circular orbit about Earth (or other start Biosphere Substrate) by some version of Yunitskiy’s General Planetary Vehicle (GPV) Orbital Launch Ring located exactly on & along the entire Equator of Earth (or any other start Biosphere Substrate) [Yunitskiy, 1982; Yunitskiy, 1987], and assembled on orbit about Earth (or other start Biosphere Substrate). Then PHBSs are delivered using ion thrusters to the near-sun orbit (slightly closer than the orbit of Mercury scale), and deployed there on stable heliocentric circular orbit, or assembled from parts delivered per multiple launches. The orbital velocity of the PHBSs is calculated lower than compensating the solar gravity for that orbit, because another force it must compensate for is the solar wind & radiation pressure on the large surface area of the light harvesting thin-foil Fresnel lens, in order to achieve almost or even exactly passive orbit maintenance. The PTSSs with some initial reservoirs of propellant enough to reach the targeted Planetesimal & provide soft landing on it, after being launched from Earth (or other start Biosphere Substrate) and assembled on orbit, deploy their 2-axis gimballed photoreceivers photovoltaic arrays on their orbits around Earth (or other start Biosphere Substrate) and being at highly irradiated orbits can start their way towards the target Planetesimal, then is found and tracked by the power beams from the PHBSs concentrators continuously during the entire mission. At least 4+ or more PHBSs equally spaced around Sun will be needed to start, to power PTSSs during entire true anomalies range, on all sides of a Planetary System host star, because there is a blind spot behind a star, and too high angles between solar radiation pressure & power beam make PHBS attitude control significantly more difficult & expensive. Although much delayed, there should be constant radio communication between each PTSS & PHBS pair, model predictive control as orbits propagations are predictable and sophisticated negative feedback control system always decreasing the mismatch error between the PTSS photoreceivers' photovoltaic array center & the PHBS arriving light beam center, determined by the power intensity gradient. The propulsion system, most likely Ion Thruster because they provide the highest exhaust velocities thus are most propellant efficient from all presently existing interplanetary propulsion systems & can use virtually any evaporated gaseous ionized material as propellant, accelerates each PTSS to each targeted Trans Neptunian Object (TNO) Planetesimal rich in N, H, & other volatiles, or other required limiting chemical elements. PHBSs' power beams track each PTSS providing it with electricity during all the entire mission, from start to end. Each PTSS can & should use all the Gravity Assists orbital maneuvers around each intermediate gas giant planet, in order to save propellant & time, because we are actually burning our payload. When reaching each targeted TNO Planetesimal, each PTSS decelerates & approaches it with decreasing velocity relative to the targeted planetesimal, & after the final deceleration, almost run out of propellant - makes soft landing & mounts on it using deployable robotic legs with hooks, & covers the entire irradiated surface of the Planetesimal with deployable 2-axis gimballed photovoltaic photoreceivers, which also decreases valuable volatiles losses via evaporation while approaching towards the star later. Then, PTSS drills the planetesimal material with the drilling rig, using the shredded planetesimal material as propellant - melted, evaporated, ionized & accelerated using the concentrated beamed solar power received. By each ion thruster engine 2-axis gimballed thrust vectoring, the PTSS helps to compensate for and stop the planetesimal rotation, in order to make the protoreceivers array oriented at the most efficient angle perpendicular to the beam, and Ion Thrusters exhaust against the Planetesimal orbital velocity vector. Then, each Planetesimal is accelerated by the PTSS mounted on it through the optimal trajectory in terms of energy to Mars, Venus, or other destination Biosphere Substrate, using all the Gravity Assists orbital maneuvers around all the intermediate gas giants to save energy thus propellant & time. When each planetesimal is approaching and put on course to collide the destined Biosphere Substrate under the required angle (impact tangential to the desired Equator, in order to adjust the planetary rotation to optimal, in most cases) - the PTSS system with the full propellant reservoir can detach from the planetesimal, & be ready for the next Planetesimal Redirection mission, or be sent to maintenance on orbit, if necessary, depending on longevity of its electrical control components achievable by that time, or if it won't survive another mission - just crush with the Planetesimal. Both PTSSs & PHBSs systems might be reusable as long as possible for all components & systems to operate reliably.

According to [Pinigin, 2000], modern technologies allow real mechanical hardware measurements & manufacturing precision of $\approx\mathbf{10^{-8}\ m}$ order of magnitude in linear dimensions. Which is, in angular dimensions, about $\approx$**0.001"** order of magnitude with 2-3 meter telescope mount length [Pinigin, 2000].

Solar wind pressure around Mercury's orbit varies a lot over time depending on solar activity [Milillo et al, 2020] and can be calculated as $\mathbf{P_{SW} = m_p{*}n{*}V^2 \approx 15\text{-}30\ nPa \approx 3{*}10^{-8}\ Pa}$, let's take the worst case scenario for calculations [Diego et al, 2020]. The sunlight pressure around Mercury's orbit is $\mathbf{P_{SL} = 2{*}(G_{SC}/(c{*}R^2)){*}(\cos(\alpha))^2 \approx 5.97{*}10^{-5}\ Pa}$ [Williams, 2007; Kopp & Lean, 2011]. The total dynamic pressure of sunlight and solar wing combined around Mercury's orbit is $\mathbf{P_{dyn}} = \mathbf{P_{SW}} + \mathbf{P_{SL}} \approx \mathbf{5.97{*}10^{-5}\ Pa + 3{*}10^{-8}\ Pa \approx 5.973{*}10^{-5}\ Pa}$. It is important to note that both dynamic pressure and specific solar power per unit area are inverse proportional to the distance from a star squared, $\mathbf{P_{dyn} \propto 1/R^2}$ where R is planetary circumstellar orbit radius or semi-major axis or a spacecraft distance from the star, or

$$\mathbf{P_{dyn1} / P_{dyn2} \propto R_2^2 / R_1^2} \quad (1)$$

For the simplest rough estimation of the order of magnitude, we might assume that the orbital Fresnel lens dominates the mass of a PHBS and all its other parts are less than order of magnitude of the orbital Fresnel lens mass thus might be neglected on this scales comparison, which might not always be true, but can help for simplified rough evaluation of the order of magnitude. We can assume that the PHBS orbital Fresnel lens is made of ultra-thin ultra-lightweight lucid nano-foil concentric rings and has a skeleton and areal mass comparable to modern solar sails, although both assumptions might be significantly reconsidered with the technologies development and the systems parameters further clarification.

If a modern lightsail aerial density is around $\mathbf{\rho_A \approx 0.031kg/m^2}$ [Negri & Biggs, 2019; Gong et al, 2019]. Then, the sunlight pressure acceleration on a modern solar sail around Mercury's orbit [Tresaco et al, 2016; Trofimov et al, 2018] will be:

$$\mathbf{a_p = P_{dyn}/\rho_A \approx (5.973*10^{-5}kg/s^2m)/(0.031kg/m^2) \approx 0.0019m/s^2}. \quad (2)$$

The average Mercury's heliocentric orbital velocity is $\mathbf{\approx 4.74*10^4\ m/s}$, with the average orbit radius of $\mathbf{\approx 5.79*10^{10}\ m}$ [Williams, 2007]. Thus, the centrifugal acceleration of spacecraft is almost equal to the acceleration of the Sun's gravitation at that orbit, will be:

$$\mathbf{a_c = v_{PHBS}^2/R_{PHBS} \approx (4.74*10^4\ m/s)^2 / 5.79*10^{10}\ m \approx 0.0388\ m/s^2}. \quad (3)$$

As for Terraforming we will need to do structures of maximal scale, we can't use rare elements like GaAs, and may use only cheap abundant Si based photovoltaic elements, as Si chemical element is so abundant on Earth that allows us to scale up power systems as much as we need. Photovoltaic arrays made from Silicon single-crystal non-concentrator have practically demonstrated up to **26.1%** efficiency [https://www.nrel.gov/pv/cell-efficiency.html].

In order to compare the thermonuclear rocket propulsion system proposed by Dr. Zuibrin in 1993 vs solar-powered deep space electric propulsion systems for Planetesimals Redirection for Terraforming purposes proposed in this work, as a reference size we can take the same average Planetesimal mass and diameter as in Dr. Zubrin's publication on Planetesimals Redirection Terraforming [Zubrin et al, 1993]: 10 billion tons = $\mathbf{10^{13}}$ **kg**, **2.6 km = 2600 m** diameter or **1300 m** radius Planetesimal, and assume the same useful power for each propulsion system of **20 GW**. But the target Planetesimal orbit semi-major axis should be not 12 AU where very few if any such Planetesimals can be found, but at least 30 AU or farther, at **30-50 AU** Kuiper Belt where volatile rich Planetesimals are confirmed to be abundant according to modern observation data [Jewit et al, 1998, 2001, 2004; Delsani et al, 2006]. That was unknown in 1993 when Dr. Zubrin published "Technological requirements for Terraforming Mars" so his assumptions can be considered scientifically fair due to lack of data at that time, which became known during the recent decades thanks to the launch of modern orbital telescopes doing Exoplanets detection & characterization, astrometry and other surveys, and giving valuable info about Planetesimals on our exterior far outer orbits often as by-product.

## 3. Theory & calculations

Regarding Planetary surface chemical composition deficiencies scales for Terraforming.

For Mars Terraforming. Mass of Earth's atmosphere is about $\mathbf{M_{atmE} \approx 5.148*10^{18}\ kg}$. Earth to Mars surface area ratio is:

$$\mathbf{S_{pE}/S_{pM} = 510\ 064\ 472\ km^2 / 144\ 371\ 390\ km^2} = *3.533 \text{ times}. \quad (4)$$

Or, $\mathbf{S_{pM}/S_{pE} = *0.283}$ times or **28.3%**, and Mars gravity 0.3728 $m/s^2$ is about **0.38 g** or **38%** of Earth gravity. Atmospheric pressure is force exerted by atmosphere weight on surface area, **P = M*g/S**, or **M = (P*S)/g**, thus $\mathbf{M_{atmM}/M_{atmE} = 0.283/0.38 = 0.7447}$. So, to make Earth atmosphere pressure on Mars will require $\mathbf{M_{atmM} \approx 5.148*10^{18}\ kg * 0.7447 = 3.8339*10^{18}\ kg}$ of total atmosphere mass. As melting everything on Mars will give about 7% of Earth atmospheric pressure, 93% of that atmosphere mass, or $\mathbf{3.565527*10^{18}\ kg}$ has to be imported. Volatiles can be about half of the Planetesimals' mass, $\mathbf{5*10^{12}\ kg}$ for each reference Planetesimal, and the rest mass rocky cores, so we can consider roughly $\mathbf{7.131054*10^{18}\ kg}$ of volatile rich TNO Planetesimals mass is required to be imported by Solar-powered Planetesimals Redirection operations for Terraforming to make the same pressure as Earth breathable atmosphere on Mars. This is $\mathbf{N = M_{atm}/m_{pv} = 3.565527*10^{18}\ kg\ /\ 5*10^{12}\ kg = 713106}$ of the reference size Planetesimals to import breathable atmosphere! We also need to import at least minimal hydrosphere for agriculture. Average of 1 m water precipitation on all Mars surface might be around bare minimum for growing food [Zubrin et al, 1993], but optimal Terraforming quality for maximal carrying capacity & productivity requires much more water, a lot of water will be absorbed by soil and go into underground caverns & voids, so 1 m water layer total precipitation is arid dry Mars. The reference Planetesimal volume is $(4\div3)\times\pi\times(2600m)^3$ = $\mathbf{7.3622\times10^{10}\ m^3}$. Mars radius $r_m$ = 3396.2±0.1 km = 3396200 m. Mars surface area is $S_{pM} = 4\times\pi\times r_M^2 = 4\times\pi\times(3396200m)^2$ = $\mathbf{1.449427107\times10^{14}\ m^2}$. Covering all Mars' surface with an average of 1 m thick layer of Planetesimal material requires $V_{HM} = S_{pM} * h_H = (4\times\pi\times(3396200m)^2)\times1m$ = $1.449427107\times10^{14}\ m^3$ volume of water. This is $1.449427107\times10^{14}\ m^3\ /\ 7.3622\times10^{10}\ m^3$ = **1969** reference Planetesimals volumes. Planetesimals have not only volatiles but rocks, if we assume on average half rocks half volatiles mass of TNO Planetesimals, we might need about **1969 * 2 = 3938** reference Planetesimals to provide minimal hydrosphere to Mars. Yet this is around bare minimum to make Mars somehow habitable, for full Terraforming quality much more water will be required. Thus, the bare minimum requirement for Terraforming Mars at least somehow is mass or number of TNO Planetesimals to import atmosphere & hydrosphere **713106 + 3938 = 717044** reference size volatile rich TNO Planetesimals! Yet this is bare minimum, full Terraforming quality making Mars maximal carrying capacity will require orders of magnitude more TNO Planetesimals delivered.

For Venus Terraforming, it is harder to estimate the required Planetesimals mass to deliver the limiting chemical elements - we will need mostly Hydrogen for Venus which is the lightest element, but even if we select only the most H rich TNO Planetesimals for Venus, in order to deliver Hydrogen we will need to deliver many times more mass of other chemical elements present in any Planetesimals. Venus radius is $r_V$ = 6051.8 km = 6051800 m, surface area $S_{pV} = 4\times\pi\times r_V^2 = 4\times\pi\times(6051800m)^2$ = $\mathbf{4.602343167\times10^{14}\ m^2}$ = 460 234 317 $km^2$, close to Earth radius & surface area, is about ***3.175** times bigger than Mars surface area, almost like Earth. Venus atmosphere mass is $M_{atmV} \approx 4.8\times10^{20}$ kg out of which ≈96.5% is $CO_2$ and ≈3.5% is $N_2$, other gasses are in trace amounts in ppm [Basilevsky et al, 2003], so $CO_2$ mass in today's Venus atmosphere is $0.965*4.8\times10^{20}$ kg = $4.632\times10^{20}$ kg. Earth atmosphere mass is $M_{atmE} \approx 5.148*10^{18}$ kg out of which 21% is $O_2$ 78% is $N_2$, thus we should leave in the Terraformed Venus atmosphere all the $N_2$ which can be later fixed into soil by nitrogen fixing bacteria colonies on roots of legumes plants in the end of Terraforming after material importing is complete and biological seeding starts, and $0.21*5.148*10^{18}$ kg ≈ $1.081*10^{18}$ kg of $O_2$. Thus, $4.632\times10^{20}$kg-((12+16+16)/(16+16))*$1.081*10^{18}$kg = $4.617*10^{20}$ kg of $CO_2$ has to be converted into graphite & water via the Bosch reaction [Birch, 1991]. Because C has atomic mass 12, N has atomic mass 14, O has atomic mass 16. Bosch reaction is $CO_2(g) + 2\ H_2(g) \rightarrow C(s) + 2\ H_2O(l)$, with Fe as catalyst, it requires 4 moles of H per 1 mole of $CO_2$, thus total mass of hydrogen to Terraform Venus is $m_{HV} = (4.617*10^{20})\times(4\div(12+16+16))$ = **$4.2\times10^{19}$ kg** of H to convert all required $CO_2$ into graphite or black soil, and water. And we will need to spray metal dust in relatively small amounts in upper layers of Venus atmosphere, less than an order of magnitude mass than other imported atmospheric components, to catalyze the Bosch reaction, and in order to bind toxic $SO_2$ into sulfur salts of metals [Birch, 1991] that are safe & can be used as fertilizers in agriculture. $SO_2$ in Venus atmosphere is 150 ppm [Basilevsky et al, 2003], which is $0.00015*4.8\times10^{20}$kg = **$7.2*10^{16}$ kg**. This is 3 orders of magnitude less that the mass of Hydrogen required to be imported, thus mass of Fe or other metals or alkaline elements to bind $SO_2$ into salts of metals that can be used as fertilizers in agriculture, can be neglected in a first order of magnitude analysis compared to the mass of H needed to be imported for at least minimal habitability of Venus. If Planetesimals were made of pure H, it would require $4.2*10^{19}$kg/$10^{13}$kg = **4 200 000** of the reference size Planetesimals, but no Planetesimal can even have a big fraction of H - in order to deliver some H by Solar-powered Planetesimals Redirection for Terraforming operations, many orders of magnitude or at least many times more mass will be required to be delivered, as chemical separation is very complex and costly process requiring heavy expensive equipment. Another reason for Solar-powered Planetesimals Redirection for Terraforming needed to Terraform Venus is to correct Venus rotation thus making daylength & tilt similar to Earth by striking the delivered Planetesimals with high velocity tangentical to the desired Planetary equator [Birch, 1991], that will spin a Planet as needed. The total amount of material needed to be delivered to Terraform Venus can be much decreased if we somehow manage to energy efficiently deliver Hydrogen rich atmosphere from Jupiter to Venus by huge balloons scooping & sealing the gas giant atmosphere, using it as propellant to deliver most H to Venus, but such huge balloons will also require significant mass & manufacturing efforts, this is a topic for a separate article.

For ExoPlanets Biosphere Substrates Terraforming. How can we estimate the average amount of volatile rich remote Planetesimal material to Terraform at least 1 average ExoPlanet Biosphere Substrate inside Circumstellar Habitable Zone in a new Planetary System? We have so little data on ExoPlanet Biosphere Substrates atmospheres [Deeg et al, 2018; Tinetti et al, 2018; Wordsworth, 2022; Zi-xin & Jiang-hui, 2024; Fu et al, 2025] that the only fair answer is we don't know. But for a first order of magnitude estimate or at least educated guess, let's take the bare minimal average amount of material required to be delivered for the 3 Planets inside the Solar System Circumstellar Habitable Zone for Terraforming and divide it by the number of Planets to get the average. From above, **717044+** reference size volatile rich TNO Planetesimals for Terraforming Mars, **0** for Earth, and Hydrogen mass equivalent to **4200000+** reference Planetesimals for Venus, gives about (717044+0+4200000)/3 = **1639014** of the reference size Planetesimals for Terraforming an average ExoPlanet Biosphere Substrate inside a Circumstellar Habitable Zone. Venus Terraforming will require in total delivering many times more mass than that, but let's assume we always first Terraform the easiest ExoPlanet Biosphere Substrate in each new Planetary System, and after inhibiting it we build Yunitskiy's GPV Orbital Launch Ring on the Equator and Terraform all other Biosphere Substrates in each ExoPlanetary System.

Yet, the amount of all the nuclear fuels discovered on Earth [Hofmann, 1988] is rather little compared to the scale of the energy required for providing all Biosphere Substrates' surface chemical composition including atmosphere, hydrosphere & soil, rotation parameters including daylength & tilt, and planetary magnetic fields close to optimal for our posterity maximal reproduction in long-term [Morozov et al, 2021a]. According to WNA [WNA, 2025], there are about $6*10^9$ kg of known Uranium yet available for mining left on Earth, which contains $7.939*10^{13}$ J/kg * $6*10^9$ kg = $4.7634*10^{23}$ J total energy, out of which with 33% nuclear reactor efficiency only about **$1.571922*10^{23}$ J** can be realistically converted into useful electric energy. Delivery of **1639014+** Planetesimals, each on average **$10^{13}$ kg**, with total delta-v 1 km/s = 1000 m/s for each on average, will take planetesimals propulsion energy of:

$$E_{PP} = \Sigma 0.5*m_i*\Delta v_i^2 = 0.5*(1639014*10^{13}\ kg)*(1000m/s)^2 = 8.19507*10^{24}\ J \qquad (5)$$

of efficiently used energy for propulsion, which is an order of magnitude more than all Uranium still available on Earth has! That is why using concentrated beamed Stellar light power as the energy source for Planetesimals Redirection operation delivering the limiting chemical elements for Terraforming is essentially necessary. **10 000** of **$10^{13}$ kg** Planetesimals delivered can provide only about 1 meter high water precipitation layer on Mars [Zubrin et al, 1993], it is about minimal Terraforming quality to make Mars at least somehow a little water available arid Planet minimally habitable for humans as significant amount of water is also absorbed into soils & sinks in underground voids, 1 m precipitation for an absolutely dry planet is far less than the amount needed for abundant water rich hydrosphere providing maximal carrying capacity possible over generations, but still an order of magnitude less than imported mass to make atmosphere required. In order to make sufficient moons for Mars & Venus to generate natural planetary magnetic fields by tidal forces, increase Mars size to near Earth size and rich in volatiles with abundant rivers and big bodies of water to grow more food, make Mars size or even Earth size Planet from dwarf planet Ceres assembling on it the Main Asteroid Belt & remote volatile rich TNO Planetesimals with also surface chemical composition comparable to Earth - would require many millions of reference size TNO Planetesimals delivered, definitely many orders of magnitude more power than all nuclear fuels available on Earth contain, thus is not possible without developing Solar-powered Planetesimals Redirection for Terraforming technologies, due to scarcity of nuclear fuel (Uranium & Thorium) in Cosmos [Lodders, 2003].

What is more, the amount of nuclear fuel available might be the principal limiting factor for the final stages of far-future Interstellar Colonisation missions, determining how many Planetary Systems can we Terraform & inhabit from one Planetary System, thus our Cosmic Expansion rates, thus Uranium must be preserved for the purpose of Interstellar Cosmic Expansion of our posterity [Morozov et al, 2021a]. An Interstellar Colonization Spacecraft will need to carry humans genetically Minimal Viable Population (MVP) of at least **10000+ people** maximally genetically diverse representing all races most distant genetic lines from all continents most far from each other [Smith, 2014], to sufficiently decrease chances of genetic degradation & extinction because of inbreeding, to provide enough genetic diversity to mitigate inbreeding effects over dozens of generations.

Highly closed Bio Technical Life Support System (BTLSS) regenerating atmosphere, water & food was estimated to be at least 1500 kg per human for 3 years [Bartsev et al, 2003], for generations BTLSS requires at least **2000+ kg** BTLSS mass per human. Such BTLSS includes LED irradiated shelves with edible plants, some species growing on hydroponics where solutions are prepared by human exometabolites alternating electric current activated “wet incineration” in $H_2O_2$ solution [Trifonov, 2012; Trifonov et al, 2015; Tikhomirova et al, 2019] synthesized inside BTLSS via electrolysis [Kolyagin et al, 2013], and part growing on Soil Like Substrate (SLS) which is rapid composting of grinded shredded straw and other planet wastes using special bacteria, fungi, and worms [Morozov et al, 2018b; Morozov, 2021b]. Gas products processing methods for above mentioned waste treatment recycling processes are developed and their acceptable safety for BTLSS is confirmed [Sutormina et al, 2011; Trifinov, 2012; Tikhomirov et al, 2012; Trifonov et al, 2020]. Edible plants produce oxygen for breathable atmosphere, food, and clean water transpirated and condensed on cold tubes, then drinking water mineralized and other used as-is [Gitelson et al, 2003]. Such BTLSS will be performed as a torus rotating about its center of mass probably with the entire Interstellar Spacecraft to provide artificial gravity, as humans cannot function long without gravity, and can't restore functionality after more than a year in microgravity [Charles, 2016]. The volume of the BIOS-3 experimental facility is about 315 $m^3$ [Gitelson et al, 2003], which provided oxygen, water and food for up to 4 people living in it. Thus, 315 / 4 = **79 $m^3$** of ecosystem volume per human approximately is required to cover all life-support needs & provide a workplace during several generations. To give a very rough order of magnitude estimate, ISS living volume available for humans is 916 $m^3$, total mass 417 000 kg [ISS, 2016]. Thus, 916 $m^3$ / 417 t is about 2.2 $m^3$ of living volume per 1000 kg of total ISS mass. Thus, ≈ 79 $m^3$ / (2.2 $m^3$/t) + 2 t ≈ 38 t ≈ **38 000 kg** mass of spacecraft hull & BTLSS per human. Surely the ISS has a lot of equipment we don't need in Interstellar Spacecraft, but also we will need to take seeds of trees & all kinds of useful plants, fish & animals to seed each Terraformed ExoPlanet, who can be part of BTLSS, & many other equipment, so we can use rough order of magnitude estimate of dozens tons per human with spacecraft hull & BTLSS.

Regarding the Terraforming Equipment mass to Terraform & inhabit at least one ExoPlanet - N.E.R.V.A. rocket engine dry mass was about 34 t = **34000 kg** [http://www.daviddarling.info/encyclopedia/N/NERVA.html] to deliver Planetesimals to Terraform at least one ExoPlanet to start life in a new Planetary System. In Dr. Zubrin's concept [Zubrin et al, 1993] 4 N.E.R.V.A. type NTR Engines were used to deliver a **$10^{13}$ kg** Planetesimal to Mars. 34 000 kg * 4 = 136 000 kg, and let's leave 64 000 kg for drilling rig & propellant feed & propellant tanks & thrust vectoring & landing grips & controls & other systems, so about **≈200000 kg** dry mass to deliver a Planetesimal using nuclear propulsion. Let's assume Interstellar Colonisation Spacecraft with **10000 people** in highly closed BTLSS and Terraforming equipment is releasing nuclear Planetesimal Tugs on outer orbits before reaching target Planets, use all possible gravity assist maneuvers to decrease circumstellar speed, and it takes only about **≈80000 kg** propellant on average for each nuclear Planetesimal Tug, thus mass of each in loaded Starship is about **280000 kg**. Other Terraforming Equipment is orders of magnitude less mass than the Planetesimal Tug Spacecraft Systems, Terraforming a Planet is mainly delivering enough limiting chemical elements by Planetesimals Redirection operations to provide acceptable surface chemical composition for our posterity, assume at least one ExoPlanet in every Planetary System we want to inhabit is inside Circumstellar Habitable Zone so we don't need starlight orbital concentrators or dissipators, and biological seeding is the easiest & cheapest part of Terraforming, moon if needed we can assemble with Solar-powered Planetesimals Redirection for Terraforming after delivering initial atmosphere & hydrosphere & starting to inhabit the ExoPlanet to start using its material to produce Spacecrafts.

We have very little data on ExoPlanetary Systems chemical compositions yet [Deeg et al, 2018; Tinetti et al, 2018; Wordsworth, 2022; Zi-xin & Jiang-hui, 2024], but we already know that most ExoPlanets have surface chemical composition far from optimal for life, especially around M-type stars which are about 75% Planetary Systems in the galaxy most ExoPlanets might have lost most atmosphere due to proximity of Circumstellar Habitable Zone to the star and stellar variability, so importing atmosphere and hydrosphere by Solar-powered Planetesimals Redirection operations for Terraforming might be most common, biggest scale of effort &

spacecraft payload mass, & most important work in Terraforming & inhabiting ExoPlanets [Morozov et al, 2016]. At present stage of ExoPlanetary Science development, it is not yet possible to give accurate estimate of average ExoPlanets surface chemical composition in our stellar cluster, so let's extrapolate from Solar System 3 Planets inside Circumstellar Habitable Zone, and assume that it will take on average delivering about **≈1639014+** of $\mathbf{10^{13}}$ **kg** volatile rich Planetesimals to minimally Terraform at least one easiest Planet in a different Planetary System. Thus, 10000 crew * 38 000 kg per human = $\mathbf{3.8*10^{8}}$ **kg** for humans MVP with BTLSS & living space and 1639014 * 280000 kg = $\mathbf{4.589*10^{11}}$ **kg** Terraforming equipment (three orders of magnitude more mass than humans with BTLSS & structures & other equipment) to Terraform & inhabit at least one ExoPlanet in average nearby Planetary System which for sure differs a lot between Planetary Systems, so $\mathbf{4.593*10^{11}}$ **kg** payload mass on average to Terraform at least one easiest ExoPlanet and start human population reproduction in a new Planetary System. So, for very rough order of magnitude estimate, we can assume $\mathbf{4.593*10^{11}}$ **kg** payload mass. Interstellar Flight requires acceleration & deceleration, $2*0.5*m*v^2$ = $\mathbf{m*v^2}$, thus:

$$\mathbf{E_{PS} = m*v^2 = 4.593*10^{11}\ kg * (0.1 * 299\ 792\ 458\ m/s)^2 = 4.12798*10^{26}\ J} \qquad (6)$$

Starship propulsion energy required to accelerate payload of such an interstellar spacecraft to 10% speed of light and decelerate it to interplanetary speeds, which is three orders of magnitude more energy than all nuclear fuel available to mine left on Earth can give!

Thus, Interstellar Colonization Spacecrafts are impractical on nuclear propulsion due to scarcity of Uranium and Thorium in Cosmos [Lodders, 2003]. Thus Starships must be Electro-Magnetic Sailships with Electric Sail hundreds kilometers or more in diameter accelerated by particles beam relativistic accelerator [Landis, 2001] from departure Planetary System that will take about entire surface of Mercury covered with Silicon based photovoltaic arrays to power [Morozov et al, 2016], and powering life support system by laser [Forward, 1984; Frisbee, 2004] during acceleration phase which is less than 10% of the way length [Morozov et al, 2016], then after achieving maximal velocity break decelerate with Magnetic Sail via interstellar magnetic fields & generate eclectic power [Andrews & Zubrin, 1990; Matloff & Johnson, 2005] for BTLSS most of the way, but eclectic power MagSail generates is proportional to the spacecraft velocity squared so can't provide enough power for humans to survive when too low velocity to gain enough power but too far from target Planetary System to use starlight concentrators with fiber optics & photovoltaic panels, so on final stages of Interstellar Colonization missions when too far from both Planetary Systems, nuclear fuel might be irreplaceable to keep humans alive [Morozov et al, 2016], as BTLSS BIOS-3 had total power consumption 240 kW or **60000 W** per crew member [Gitelson et al, 2003], which can be decreased to 30000 W per crew member with modern more efficient LED technologies. So, with life support costs of **30000 W** per crew member, MVP of **10000 people** requires total power for life support:

$$\mathbf{P_{MVP} = P_h*n = 10000\ people * 30000\ W/human = 3*10^8\ W =} \qquad (7)$$
$$= 356.25d/yr * 24h/d * 60m/h * 60s/m * \mathbf{3*10^8\ W = 9.46728*10^{15}\ J/year}$$

So $\mathbf{9.46728*10^{15}}$ **J** energy every year is required to keep MVP alive when too far from any star to use starlight power for just regenerating air water & food to keep humans alive. It is easy to burn precious nuclear fuel, but we can't pack energy back to that density. So, we should do everything within each Planetary System using stellar light power as energy source, and preserve precious extremely limited Uranium for Interstellar Colonization purposes where it might be irreplaceable - this is another motivation to develop Solar-powered Planetesimals Redirection for Terraforming technologies, even though they are more technologically complex than NTR propulsion. Enabling the possibility of using concentrated beamed Solar power as the energy source of Planetesimals Redirection operations for Terraforming - can increase their possible scale to many orders of magnitude, as much as needed so power is not a limit anymore, but only workforce determines the scales of Terraforming quality!

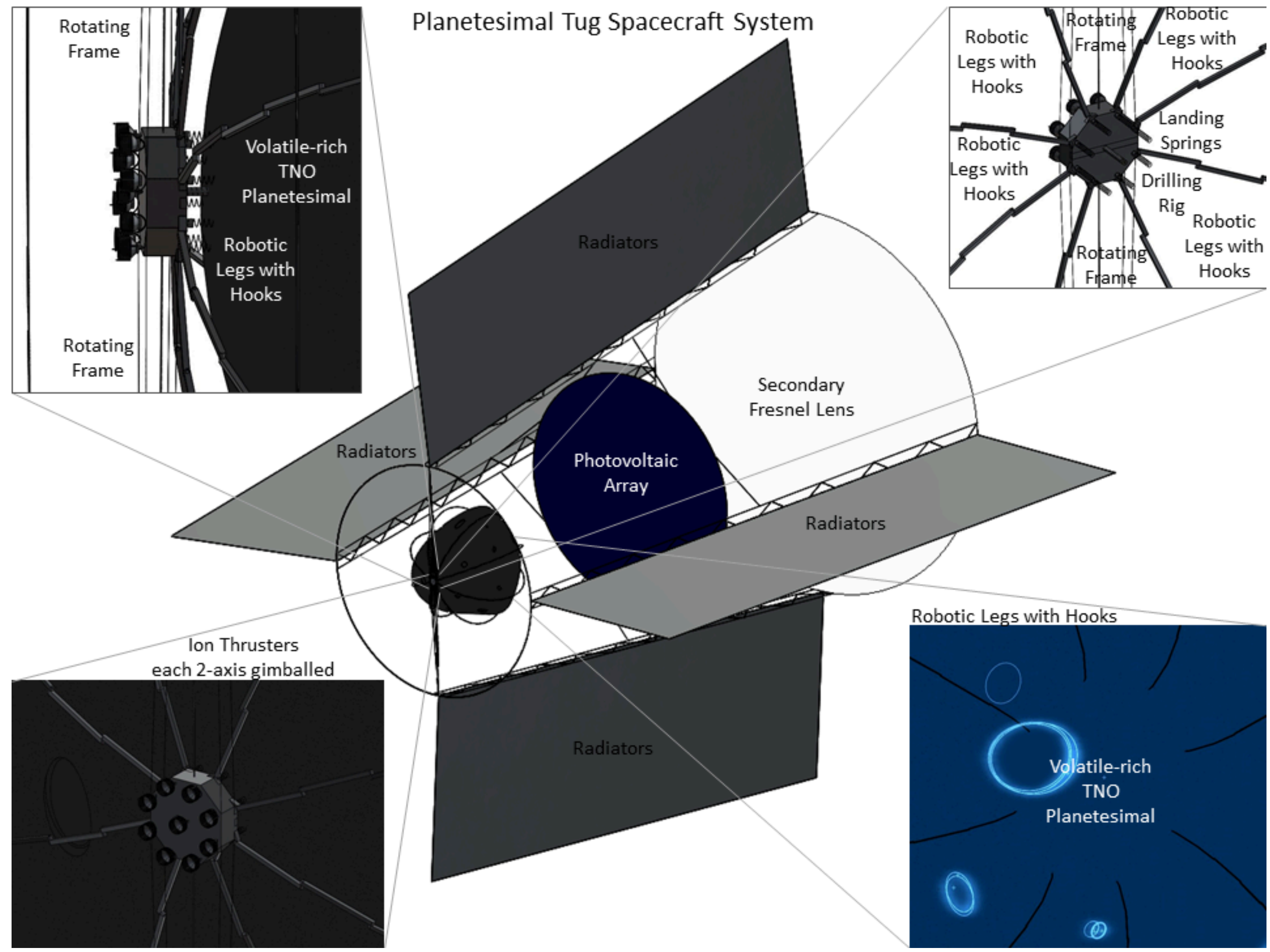

Fig. 6. Planetesimal Tug Spacecraft System preliminary design

The principal physical & technical limitations for scaling up Solar-powered Planetesimals Redirection for Terraforming operations might include:

1. Astrodynamical long-term stability of the PHBSs. Around Mercury's orbit solar wind pressure is up to $\mathbf{P_{SW} \approx 3*10^{-8}}$ **Pa** [Diego et al, 2020; Milillo et al, 2020] & solar radiation pressure is about $\mathbf{P_{SL} \approx 5.97*10^{-5}}$ **Pa** [Kopp & Lean, 2011], which give $\mathbf{P_{dyn} \approx 5.973*10^{-5}}$ **Pa**. A modern lightsail mass is $\mathbf{\approx 0.031 kg/m^2}$ [Negri & Biggs, 2019; Gong et al, 2019], orbital Fresnel lens mass will be dominated by ultra-thin ultra-lightweight nano-foil concentric rings, thus not differ much from solar sail aerial density. Then pressure acceleration on a light harvesting lens around Mercury's orbit will be $\mathbf{a_p \approx (5.973*10^{-5}\ kg/(s^2m))\ /\ (0.031\ kg/m^2) \approx 0.0019\ m/s^2}$. Corresponding stable circular heliocentric orbital velocity is about $\mathbf{\approx 4.74*10^4\ m/s}$ [Williams, 2007], centrifugal = gravity acceleration at that heliocentric orbit will be $\mathbf{a_c = v^2/R = (4.74*10^4\ m/s)^2\ /\ 5.79*10^{10}\ m \approx 0.0388\ m/s^2}$. Acceleration $\mathbf{0.0019 m/s^2}$ is a little fraction of $\mathbf{0.0388\ m/s^2}$, so (almost) passive orbit maintenance is relatively easy to compensate by proportionally (**≈5%**) slower heliocentric orbital velocity of the PHBSs.

2. Scaling up the power systems. Cheap and abundant Si based photovoltaic elements have been practically demonstrated to have efficiency of **26.1%**, and the theoretical limit of their efficiency is above **30+%** [Bhattacharya et al, 2019], there is enough $SiO_2$ on Earth to manufacture way more photovoltaic panels then we need. The only principal physical limit that we can observe is that the size of the thin-foil power harvesting orbital lens is limited by its gravitational collapse on itself, which depends on materials properties, the compression strength of the orbital Fresnel lens "skeleton". The lens size limit can be increased by compensating the gravity force collapsing it by centripetal force made by rotation of the orbital lens about its center of mass axis of symmetry, which can also be used for deployment - but this makes the beam pointing and tracking much more difficult, and hardly ever would be needed. The gravitational collapse is not a power limit anyhow, as there is no need to make one big lens - the number of lenses is almost unlimited, this problem can be overcome by producing orbital arrays of smaller lenses, deploying them in precisely oriented formation flying arrays, that have space to grow in scale up to Dysonian swarm even closer than Mercury orbit, though orders of magnitude much less power is needed for Solar-powered Planetesimals Redirection for Terraforming operations. No principal limits here.

3. Reciprocal radio communication with even much more distant spacecrafts have been not only theoretically studied and proven [Hemmati et al, 2011] but also practically demonstrated by the Voyagers spacecrafts [https://voyager.jpl.nasa.gov/mission/ ; Ludwig et al, 2002]. No principal technical limits here.

4. Scaling up the propulsion systems. Dr. Zubrin calculated [Zubrin et al, 1993] that using a quartet of **5000 MW** NTR thrusters (total **20 GW**), we can deliver a **≈2.6 km** diameter **$10^{13}$ kg** TNO Planetesimal rich in $NH_3$ & other volatiles limiting chemical elements, from the Kuiper Belt at about **30-50 AU**, where volatile rich objects are already abundant, to Mars in about **100-200 years** on energy efficient path, using all gravity assists around all the intermediate gas giant planets, & we would spend less than **8%** of the planetesimal material mass as propellant, which had exhaust velocity of only **4 km/s**! [Zubrin et al, 1993].

Ion thrusters and other modern deep space electric propulsion systems are much more propellant efficient, with exhaust velocities of **$v_{ex}$ ≈ 20–100+ km/s ($I_{sp}$ ≈ 2000–10000+ s)** [Goebel et al, 2008; Brophy et al, 2000; Choueiri, 2009; Holste et al, 2020] operated over long time, laboratory tested up to **$v_{ex}$ ≈ 150 km/s ($I_{sp}$ ≈15000 s)** [Bramanti et al 2006], theoretically unlimited up to the speed of light, but practically limited by power consumption [Goebel et al, 2008; Dale et al, 2020], thus requiring much less Planetesimal mass fraction as propellant (**<0.8%**) and can do with less delivery time, so Dr. Zubrin's estimates in 1993 are rather conservative efficiencies compared to many times more propellant efficient modern deep-space propulsion systems. Ion thrusters can theoretically use any gaseous material as propellant with more or less efficiency [Fazio et al, 2018; Holste et al, 2015; Dietz et al, 2019; Marcuccio et al, 1998]. Specific Ionization Energy for different chemical elements and molecules differ several times, less than 10 less than order of magnitude between minimum and maximum [Lenntech, 2018], water specific ionization energy is near the middle of that range [NIST, 2025], thus what propellant to use for Ion Thrusters in a question of up to several times difference in ionization power required, but not a principal feasibility issue. Most Planetesimals volatiles mass is water, and Ion Thrusters successfully operating on water as propellant have been already practically experimentally successfully demonstrated and theoretically analyzed [Ataka et al, 2021; Nakano et al, 2020; Nakamura et al, 2018; Shirasu et al, 2023]. The drilling rig can feed shredded planetesimal material into PTSS propellant tank, then stop and seal in closed volume, then evaporate it using concentrated solar power, and everything that evaporated feed into Ion Thrusters, then seal the propellant gas, open hatches outside and push the leftover solid material to sides with pistons, then seal outside hatches open drilling rig, drill, repeat the cycle again, etc.

There is no foreseen principal physical limit to scale up ion thrusters to **20 GW** & more, as even if there is - smaller deployable ion thrusters can operate as arrays where each is 2-axis gimballed for thrust vectoring, which up to all the dark side of the targeted Planetesimal can be covered with. The ion thrusters can be delivered by several launches enabled by the Yunitskiy's GPV Orbital Launch Ring [Yunitskiy, 1982; Yunitskiy, 1987], and each PTSS & PHBS can be assembled on orbit about Earth or any other departure planet if too big mass and/or too large dimensions for cargo sections of Yunitskiy's GPV Orbital Launch Ring on Equator.

The exhaust jet propulsion thrust vectoring angles will be highly controllable with 2-axis gimbaled Ion Thrusters, in order to compensate for the Planetesimal rotation around its axis of revolution, for attitude control to keep each Planetesimal with PTSS rigidly mounted on it oriented with 2-axis gimballed photovoltaic photoreceivers perpendicular towards the power beam and with the exhaust jets against the Planetesimal orbital velocity vector.

5. Attitude Control. Stopping the Planetesimal rotation about its center of mass. The PTSS mounted on the Planetesimal must be oriented with its 2-axis gimballed photoreceivers surface perpendicular towards the power beam with the angle of incidence tending to zero for maximal photovoltaic efficiency, and the ion thrusters array against the Planetesimal orbital velocity vector to propel each Planetesimal in the correct direction towards the target Biosphere Substrate in most energy efficient way. Each Ion Thruster will be 2-axis gimballed allowing rotation from almost **-90** degrees to **+90** degrees angles in both directions, thus will be able to compensate for the Planetesimals' rotation about itself. Some Planetesimals rotating too fast about their centers of mass might be less preferable candidates for delivery than those rotating slower, thus easier to stabilize at the correct attitude. Yet, even in case if the PTSS 2-axis gimballed Ion Thrusters' power would be insufficient to stop the Planetesimal rotation about itself in one turn - the Ion Thrusters can be powered off for some small part of the Planetesimal rotation about itself in one turn, the PTSS can shortly turn hibernation mode with only computers and telecommunication antennas to find & track it on, operating from accumulators batteries during some short fraction of the Planetesimal rotation about itself while 2-axis gimballed ion thrusters array and 2-axis gimballed photoreceivers are on the opposite side from where they should be. The power beam can be intentionally pointed a little bit aside to minimize volatiles evaporation when the opposite side of the Planetesimals with the PTSS is towards the beam, and then pointed back into the PTSS 2-axis gimballed photoreceivers center when photoreceivers again turned towards the power beam. Thus, even if a Planetesimal rotates too fast - it can be technically possible to stop its rotation about itself and make it stabilized in the correct attitude perfectly oriented towards the power beam even in several turns.

6. Power. PTSSs 2-axis gimballed photovoltaic arrays made from cheap & abundant Silicon allowing to scale up PTSSs as needed can have up to **26.1%** efficiency [https://www.nrel.gov/pv/cell-efficiency.html]. Assume **≈9000 W/m²** of concentrated irradiation power arrives at PTSS's photoreceivers. To provide 20 GW of electric power for Ion Thrusters, will require $\mathbf{2*10^{10}\ W\ /\ (0.264*9000\ W/m^2) = 8.4175*10^6\ m^2}$, or **1637 m = 1.637 km** radius (**3.274 km** diameter) photovoltaic array, less than the biggest already built. If any problem scaling up photoreceivers – they can be delivered & deployed as arrays pieces, then connected.

7. Soft landing a spacecraft on a planetesimal was already developed in details in the NASA ARM (Asteroid Redirect Mission) over years

[https://www.nasa.gov/content/what-is-nasa-s-asteroid-redirect-mission], and already practically demonstrated by ESA's Rosetta mission [https://rosetta.esa.int/], so this is not a principal physical & technical problem but only a question of funding or providing sufficient resources, workforce & political will.

8. Thermal. PTSSs' heat rejection system must hold $NH_3$, our most precious volatile, frozen in solid state to avoid losing it via evaporation during transportation. To minimize losing $NH_3$ & other valuable volatiles along the way - 2-axis gimballed photoreceivers cover all the irradiated side of each target Planetesimal from sunlight, radiators maintain Planetesimals' temperature below −77.73 ºC = **195.42 K**. Silicon based photovoltaic cells arrays best yet achieved efficiency is about $\eta_s$ **≈26.1% ≈0.261** [https://www.nrel.gov/pv/cell-efficiency.html]. Solar panels average reflectivity is q ≈ 0.3 ≈ 30% [Ahmed et al, 2021]. Ion Thrusters overall system efficiency is $\eta_p$ **≈52%** [Goebel et al, 2008], thus ≈48% = *0.48 of power delivered to Ion Trusters is wasted as heat. Radiators need to reject about ≈ 100% - 26.1% - 30% = 43.9% = 0.439 of power is wasted as heat in photovoltaic panels, and ≈ 0.261*0.48 = 0.125 = 12.5% of power is wasted in Propulsion. Thus PTSSs must reject as heat totally ≈ 43.9% + 12.5% = 0.564 = **56.4%** of incident solar power. In order to achieve useable electric power of **20 GW**, total incident light power must be $P_t = 2*10^{10}$W / 0.261 = **$7.576*10^{10}$ W** = **76.62 GW**. Out of which, wasted rejected heat power is $P_r = P_t$ * 0.564 = 76.62 GW * 0.564 = **43.21 GW**. Rejected heat power via irradiation can be calculated as [Heo et al, 2022]:

$$\mathbf{P_r = j*A_r = A_r*\varepsilon*\sigma*T^4}, \qquad (8)$$

where $A_r$ is radiating surface, ε is emissivity typically ≈0.9 for anodized Aluminium & most paints, also ε ≈0.9 for water & ice, $\sigma = 5.67*10^{-8} W*m^{-2}*K^{-4}$ is Stefan–Boltzmann constant, T is absolute temperature in K.

Thus required radiators surface area can be found as:

$$\mathbf{A_r = P_r/(\varepsilon*\sigma*T^4)} = 4.321*10^{10}W/(0.9*5.67*10^{-8}W*m^{-2}*K^{-4}*(195.42\ K)^4) \qquad (9)$$

= **580607528 m²** = **581 km²** of total radiating surface area.

Planetesimal of 2.6 km diameter = 1300 m radius itself has radiating surface area of $S_p$ = $4*\pi*(1.3km)^2$ = **21.237 km²** given the Planetesimal is blocked from sunlight by the PTSS photovoltaic arrays, so PTSS radiators are responsible for providing a total of **560 km²** of radiating surface area, or **280 km²** double-sided. Using ultra-thin lightweight foil radiative cooling 4 double-sided radiators, 560 km² / 8 = **70 km²** radiators area each on each truss. Let's take margin for other systems & safety around 10%, and use **77 km²** as designed double-sided radiators surface area on each quarter main truss. Good thermal insulation on bottom surface of PTSS Propellant tank and robotic legs hooks ends as we need to keep propellant warm and the Planetesimal cold. For example, **12 km** long *** 6.42 km** wide double-sided thin-foil lightweight radiator along each of the 4 PTSS main trusses symmetrically evenly distributed can be enough to keep the average Planetesimal temperature below $NH_3$ melting point. Though total radiators area needed might be less, as all power wasted as heat in Ion Thrusters & elsewhere should be maximally used to evaporate & heat the propellant for Ion Thrusters, which increases the propulsion efficiency, and hot exhausted propellant takes some fraction of wasted power as heat, removing it from the PTSS & the Planetesimal. Also the better thermal insulation we have on PTSS propellant tank bottom & legs bottom surfaces, everything that touches the Planetesimal - the higher we can allow the PTSS temperature to be while keeping entire Planetesimal below $NH_3$ freezing point, which allows to decrease radiators surface area thus the entire PTSS mass, out of which radiators are significant surface thus mass.

These estimates provide a first-order sizing, though future work should also assess sensitivity to emissivity, operating temperature, contamination effects, radiator mass, as well as the potential challenges associated with deployment and operation of structures at this scale.

Preliminary PTSS 3D CAD SolidWorks design is presented on Fig. 6.

9. Angular pointing & tracking precision limiting the effective power beaming distance. Modern telescopes provide **≈ ±0.001" ≈ ±$5*10^{-9}$ rad** angular error precision tracking remote objects for 2-3 m long telescope mount [Pinigin, 2000]. For such small angles, sin can be considered approximately equal to the angle in radians: **sin($5*10^{-9}$ rad) ≈ $5*10^{-9}$**. Thus, linear pointing & tracking error **Δr** of beam & receiver mismatch can be calculated as function of the beamer angular pointing & tracking error **Δθ** depending on interplanetary linear distance **d** as:

$$\Delta r = d*\sin(\Delta\theta) = d*\Delta\theta \qquad (10)$$

For an average Mars orbit semi-major axis of about **$2.28*10^{11}$ m** distance to PTSS photoreceivers & PHBS solar power harvesting concentrator with beaming collimator near Mercury orbit which semi-major axis is **$5.8*10^{10}$ m** on average or an order of magnitude less than Martian orbit semi-major axis, a PHBS can be in a wide range of true anomalies on the same side of the Sun as the a PTSS, so distance from Sun to PHBS can be neglected for a rough order of magnitude estimation. This gives us about **≈ ±$5*10^{-9}$ * $2.28*10^{11}$ m ≈ ±$1.14*10^{3}$ m** pointing accuracy error at average Martian orbit. For PTSS photoreceivers on average Ceres orbit of **$4.14*10^{11}$** m near the middle of the Main Asteroid Belt, this is about **≈ ±$5*10^{-9}$ * $4.14*10^{11}$ m ≈ ±$2.07*10^{3}$ m** pointing error accuracy. A ≈ **±2 km** pointing error can be enough to track a **2.6 km = 2600 m** diameter planetesimal covered by **3.274 km** diameter photoreceivers photovoltaic array with an additional **10 km** diameter secondary ultra-thin nano-foil lens carried by the PTSS.

But the Kuiper Asteroid Belt, where volatile rich Planetesimals begin, is more than an order of magnitude more distant & about two orders of magnitude more massive [Delsanti & Jewitt, 2006] than the Main Asteroid Belt. With Kuiper Belt Objects orbits semi-major axes ranges of about **30 AU** to **50 AU**, taking **40 AU ≈ $40*1.5*10^{11}$ m ≈ $6*10^{12}$ m** as average, pointing error **≈ ±$5*10^{-9}$ * $6*10^{12}$ m ≈ ±$3*10^{4}$ m**. A **±30 km** beam tracking error is several times bigger than a **10 km** diameter of a secondary PTSS lens. The inner boundary of the Oort cloud, beginning at **2000 AU**, is **2-3+** orders of magnitude more distant than modern tracking technologies precision can allow to follow a planetesimal with a focused power beam.

Yet, **≈ ±0.001" ≈ ±$5*10^{-9}$** rad angular precision tracking remote objects was the limit for a **≈2 m** long mount according to [Pinigin, 2000]. If we make the mount *10 times bigger, according to the triangle proportion we can get almost 10 times smaller or *0.1 error. So, if we make the mount **200 m** long, which is not too big for a few km diameter PHBS - with present limits of material processing technologies precision, we might be able to achieve pointing & tracking accuracy close to **≈ ±0.00001" ≈ ±$5*10^{-11}$** rad angular precision, which can reliably allow to track PTSSs photoreceivers by solar power beams in the Kuiper Belt with **±$5*10^{-11}$ * $6*10^{12}$ m ≈ 300 m ≈ ±0.3 km** pointing & tracking precision error accuracy in the middle of the Kuiper Belt, allowing to reliably deliver volatile rich KBO Planetesimals to Mars & Venus. There are no principal limits observed to produce pointing & tracking mount 200 m long and launch it with Yunitskiy's GPV [Yunitskiy, 1982; Yunitskiy, 1987] with other PHBSs & PTSSs parts from planetary surface Equator to low circumplanetary orbit, then deliver to required circumstellar orbits using 2-axis gimballed Ion Thrusters [Goebel et al, 2008].

Regarding Oort's cloud, with **200 m** long mount of PHBS pointing & tracking system, the error at **2000 AU ≈ $2000*1.5*10^{11}$ m ≈ $3*10^{14}$ m** will be **≈ ±$5*10^{-11}$ * $3*10^{14}$ m ≈ ±$1.5*10^{4}$ m**

≈ ±**15 km**. So, at Oort cloud the pointing accuracy is still too low to make the mission feasible with a **200 m** long mount of a PHBS pointing system. Theoretically it can be possible to do a PHBS mount **2000 m** long, which can be proportionate for a PHBS with lens several dozens km in diameter, though it is rather difficult but principally can be possible [Mohan et al, 2014].

10. Beam divergence, assuming diffraction limited optics, can be calculated as: **d=2.4*s*λ/D**, where s = 50 AU = 50*1.4959787×$10^{11}$ = **7.47989353×$10^{12}$m**, D is laser aperture diameter, laser λ = Sun peak wavelength about 500 nm = **5*$10^{-7}$m**, d = ⌀**10 km** secondary lens is the spot on the remote object receiving 84% power [Forward, 1984]. **D = d/(2.4*s*λ) = 898 m**. As explained by Dr. Forward, final aperture of the laser transmitter subsystem does not have to be a solid lens - it can be a phased array of lasers or thin-film holographic or Fresnel lens [Forward, 1984; Frisbee, 2004], Fresnel lens most preferable.

11. Orbital Mechanics of the PTSSs. Using all the gravity assists maneuvers potential about all the intermediate gas giant planets, ΔV required to propel KBO Planetesimals onto a collision course with Mars can be less than **0.5 km/s**. For **30-50 AU** semi-major axis KBO Planetesimals, energy efficient flightpath delivery time to Mars is about **100-200 years**, as calculated by Dr. Zubrin [Zubrin et al, 1993].

12. Scales of Equipment. PTSS mass is roughly about one millionth (1 ppm), **$10^{-6}$ = 0.000001** mass fraction order of magnitude of its target Planetesimal mass that it delivers from Kuiper Belt to Mars. We are going to provide more detailed PTSS design & mass calculations in next publications. Yet, for a rough order of magnitude estimate - 5 GW N.E.R.V.A. NTR engine had dry mass of 34 t = 34000 kg, 4 of them (20 GW) had mass of 136000 kg, assume 64000 kg for drilling rig, propellant tank, feed systems, thrust vectoring 2-axis gimbals, controls, and all the rest auxiliary equipment, gives roughly ≈**200000 kg** dry mass of a nuclear 20 GW Planetesimals Tug delivering **$10^{13}$ kg** Planetesimal from Kuiper Belt to Mars, or **2*$10^{-8}$** of its mass fraction, yet radiators mass of NTRs were not accounted for, lacking which can cause big loss of payload volatiles evaporating along the way, thus with radiators to keep $NH_4$ frozen mass can increase 1-2 orders of magnitude and become close to solar-powered ion thrusters propelled PTSS… 2-axis gimballed Ion Thrusters Propulsion Systems themselves are much lighter weight for the same power levels than NTRs, but photovoltaic arrays & radiators to keep the Planetesimal temperature below $NH_3$ freezing point, took significant mass that is be also needed to keep volatiles frozen but was not accounted for a nuclear propulsion system. So Solar-powered 2-axis gimballed Ion Thrusters based PTSS has mass rough order of magnitude estimate is about one millionth ≈**$10^{-6}$** of mass of the Planetesimal that it delivers from Kuiper Belt to Mars, but more exact calculations we are going to provide in the next publication about the PTSS design, as we need to complete FEA in SolidWorks to do the design mass optimization.

Free electron solar-pumped laser theoretical efficiency is up to 50% = **0.5** [Forward, 1984], practically already achieved more than >10% or **>0.1** solar powered sun-to-laser overall efficiency [Johnson et al, 2013]. Irradiation at Mercury's orbit is about ≈9126.6 W/$m^2$ [Williams, 2007], PHBS surface area about ≈7.576*$10^{10}$ W / (0.84*0.5*9126.6 W/$m^2$) = 19 764 310 $m^2$ = **19.76 $km^2$**, lens diameter ⌀≈5 km. Modern lightsails have an aerial density of ≈0.031kg/$m^2$ [Negri & Biggs, 2019], which allows to estimate PHBS mass about 19 764 310 $m^2$ * 0.031kg/$m^2$ ≈ 612 694 kg, or **≈6*$10^{-8}$** of its target delivered Planetesimal mass. Many TNO Planetesimals observed with bulk density as low as ≈1.14 g/$cm^3$ suggest significant number of exemplars with more than half mass water, ammonia & other volatiles ices exist [Vilenius et al, 2014], so if we use 0.8% of Planetesimal mass as propellant for 2-axis gimballed Ion Thrusters it is an order of magnitude less than the payload of volatile chemical elements delivered to Mars, Venus, & other

Biosphere Substrates during Terraforming.

13. Scales of Planetesimals payload mass delivered for Terraforming. As explained in the beginning of the Theory Section, at least **717044+** of the reference size Planetesimals will be required to make a breathable atmosphere of the same pressure as on Earth on Mars. And Hydrogen mass equivalent to **4200000+** reference size Planetesimals minimum is required to Terraform Venus, but in order to deliver that amount of hydrogen - many times more if not orders of magnitude more mass of other Planetesimals chemical elements will have to be delivered with it. Yet for the case of Mars that is poor minimal Terraforming quality. Making a significant size moon for Mars to generate sufficient Planetary Magnetic Field [Hossain et al, 2015] will require orders of magnitude more Planetesimals than only atmosphere delivery.

Regarding the Planetary Magnetic Fields. Mars orbit semi-major axis is $\mathbf{R_{PM}}$ = **1.52368055 AU**, thus according to the inverse square law Eq. (1) aerial density of sunlight radiation & solar wind at Mars orbit is about $\mathbf{P_{dynPM} / P_{dynPE} = R_{PE}^2 / R_{PM}^2}$ = $(1AU/1.52368055AU)^2$ = **0.431** fraction of that at Earth's orbit, thus to have same level of protection against atmosphere loss and radiation, Mars needs about **0.431** fraction of Earth planetary magnetic field strength. Mars mass is about $\mathbf{M_{PM} = 6.4171\times10^{23}}$ **kg** [Konopliv et al, 2011], Phobos mass is about $1.060\times10^{16}$ kg, Deimos mass is about $1.51\times10^{15}$ kg, so their total mass is about $(6.4171*10^{23}kg)/(1.060*10^{16}kg+1.51*10^{15}kg)$ = $\mathbf{1.9*10^{-8}}$ fraction of Mars mass. Earth's Moon is roughly 1% of 0.01 fraction of Earth's mass. Considering **0.431** fraction of Earth's sunlight radiation and solar wind intensity at Mars's orbit, a moon of 0.01*0.431 = 0.0043 fraction of Mars mass will have mass of about $0.0043*6.4171*10^{23}$ kg = $2.7594\times10^{21}$ kg or **275935300** of the reference size Planetesimals - though small units of km size volatile-rich TNO Planetesimals we should use for importing Atmospheres and Hydrospheres to Biosphere Substrates, and for making moons we can use biggest hundreds km size dense rocky Planetesimals poor in volatiles, as chemical composition of moon doesn't matter thus should be made of whatever leftover least valuable materials - it only has to provide tidal forces to strengthen current conducting fluid flows inside planetary molten cores to strengthen Planetary Magnetic Fields enough for atmosphere retention & humans reproduction. Yet we don't know how much of Earth's magnetic field strength is enough for humans reproduction & atmosphere retention - it might be 10% or 1% enough, and closer orbiting moon might require less mass to create same tidal forces & magnetic field, thus Martian moon might be 0.001 or 0.0001 fraction of Mars's mass sufficient, so it might be not 2 orders of magnitude more but comparable mass as Terraforming Venus requires - yet clearly $10^{-8}$ of planetary mass moon size is insufficient. Passive methods of Planetary Magnetic Field strengthening such as assembling a moon from leftover planetesimal material [Hossain et al, 2015] via Solar-powered Planetesimals Redirection operations for Terraforming is preferred over active methods such as huge solar-powered solenoids along entire Equator of a Biosphere Substrate, because active methods of Planetary Magnetic Field strengthening take too big fraction of most valuable Equatorial latitudes surface area for solar panels for their constant maintenance [Hue et al, 2014], that can be used for photosynthesis thus growing and feeding significantly more population thus workforce over the remaining Planetary System lifecycle if planetary magnetic fields are generated by passive methods only or as much as possible [Hossain et al, 2014] which take no surface area and no constantly occupied surface to constantly maintain them. Thus even though passive magnetic field generating methods which mean assembling a satellite might be more expensive to build, it will pay off in decreased maintenance and increase planetary population thus workforce over the remaining Planetary System lifecycle.

The bigger Mars becomes by Terraforming works the better both for photosynthesizing surface area and gravity, the closer we make Mars to Earth size & surface chemical composition the better. Making Mars or even Earth size Planet from dwarf planet Ceres and the Main Asteroid Belt will require many trillions of the reference size Planetesimals delivered. Or, millions of much bigger Planetesimals. The more Planetesimals we bring to Terraformed Biosphere Substrates the better: Terraforming quality = planetary carrying capacity is proportionate to the work done, it is hard to estimate exactly yet because Planetesimals are significantly diverse & different, but delivery of at least **millions+** of reference size volatile rich TNO Planetesimals to Mars & Venus is required for minimally acceptable Terraforming quality.

So, all technologies principally exist to the best of our knowledge to make Solar-powered Planetesimals Redirection for Terraforming concept feasible. Delivering volatile rich Planetesimals from Kuiper Belt to Mars & Venus can be feasible with present day limits of mechanical manufacturing precision for 200+ m long pointing & tracking system mount for units of kilometers diameter PHBS. While delivering Planetesimals from Oort cloud will require to increase the angular precision of tracking remote objects with a power beam most probably by increasing principal precision of mechanical parts manufacturing by **1-3+** orders of magnitude, or by making PHBSs very big starting from dozens kilometers in diameter or more. Telescope pointing & tracking precision was increased by $\mathbf{10^5}$ or error decreased by 5 orders of magnitude during recent few centuries [Pinigin, 2000], thus looks principally physically technically doable.

If the reviewers & readers see any other possible principal physical & technical limitations of the concept feasibility - we are also ready to research those.

## 4. Results

Summarising the above, Venus & Mars & most ExoPlanets can NOT be Terraformed with resources only present on them - the main work in Terraforming, requiring bigger payload mass & work efforts, that costs orders of magnitude more than all other Terraforming operations, processes, & equipment combined - is delivering the limiting chemical elements of planetary scales by Solar-powered Planetesimals Redirection operations for Terraforming. Nuclear explosions or only heating in any amount will NOT make Mars habitable, but only waste uselessly precious extremely limited Uranium necessary & probably irreplaceable for far future Interstellar Colonization which is an unforgivable crime, because Mars totally has more than an order of magnitude less volatiles than needed to make breathable atmosphere, thus importing the limiting chemical elements in planetary scales is the key, biggest & most important work for Terraforming. Spending same amount of energy on Planetesimals Redirection operations for Terraforming will produce orders of magnitude more waste heat just as a side effect then if same amount of energy was applied directly to heat Mars, but most importantly Planetesimals Redirection operations for Terraforming will deliver the limiting chemical elements. The amount of the limiting chemical element is the main, most valuable, expensive & important factor & work in Terraforming = increasing Planetary Carrying Capacity, that to most extent determines each Biosphere Substrate saturation population over the remaining Planetary System lifecycle. Thus, Solar-powered Planetesimals Redirection for Terraforming operations give many orders of magnitude more value in the long-term than their total cost is.

There is more than enough material on outer orbits of Planetary Systems to make Earth size & surface chemical composition on every Biosphere Substrate - the main principle factor determining available Terraforming quality is the amount of energy required for Planetesimals Redirection operations. Nuclear fuel is scarce in Cosmos, and all nuclear fuel available for mining on Earth is insufficient to provide adequate Terraforming quality such as breathable atmosphere & minimal hydrosphere for Mars & Venus, let alone grow Mars to Earth size & make habitable Mars-size or even Earth-size Planet from dwarf planet Ceres & Main Asteroid Belt. Thus, development of Solar-powered Planetesimals Redirection for Terraforming technologies is essentially necessary to make an Earth like environment on Mars & Venus, let alone provide optimal Terraforming quality for Mars, Venus & assemble Planet from Ceres. Nuclear fuel must be saved for far future Interstellar Colonizations missions where it might be irreplaceable, and everything inside each Planetary System we should do using host star light power, that flows the same way regardless of the extent how much we use it or not.

It is generally physically & technically principally possible to deliver volatile rich TNO Planetesimals by PTSSs using concentrated beamed solar power powering 2-axis gimballed Ion Thrusters using in-situ evaporated Planetesimal material as propellant as-is from Kuiper Belt in near future with presently available materials & technological principles & processes. From Oort cloud not yet but may be feasible with reasonable increase in mechanical parts manufacturing precision [Zhang et al, 2019] in some generations from now in a more distant future.

Power beam pointing & tracking angular precision can be increased by making the whole pointing & tracking system proportionally bigger, dividing the relative angular pointing & tracking accuracy error by how much relatively we scaled up the entire system - we can decrease the angular pointing & tracking error by 1-2-3 orders of magnitude by making the whole pointing & tracking system 1-2-3 orders of magnitude proportionally bigger than modern telescope mounts. This allows it to reliably track PTSSs' photoreceivers in the whole Kuiper belt with PHBSs' light beams powering PTSSs' 2-axis gimballed Ion Thrusters & other systems during the entire mission to deliver volatile rich TNO Planetesimals to Mars & Venus. Obtaining Planetesimals from Oort cloud for Solar-powered Planetesimals Redirection for Terraforming might still require **1-3+** orders of magnitude increase in principally available manufacturing precision of mechanical components, otherwise PHBSs might be required to be too big.

Comparison of key performance parameters of nuclear thermal rocket vs solar powered ion thrusters propulsion systems for Planetesimals Redirection operations for Terraforming is presented in Table 1:

| Parameter | NTR | Solar-powered Ion Thrusters |
| --- | --- | --- |
| Jet Thrust Exhaust Velocity | **4000 m/s** | **40 000 m/s** achieved,<br>up to 299 792 458 m/s<br>theoretically possible |
| Fraction of Planetesimal mass used as propellant to deliver from Kuiper Belt to Mars | **8%** | **0.8%** |
| Energy Budget | **$1.571922*10^{23}$ J** | Practically **unlimited**,<br>many orders of magnitude<br>more than needed |

Table 1. Comparing Nuclear Thermal Rocket vs Solar-powered Ion Thrusters propulsion systems for Planetesimals Redirection operations for Terraforming

**5. Discussion**

Large-scale infrastructure for Solar-powered Planetesimals Redirection for Terraforming follows the economy of scales, and can be launched with some variation of Yunitskiy's General Planetary Vehicle, a telescopic Orbital Launch Ring around all the entire Equator of this and every other Terraformed & inhabited Biosphere Substrate, that can provide about 5 orders of magnitude cheaper orbital launch in terms of resources & energy objective physical parameters than modern chemical rocket launch systems can achieve [Yunitskiy, 1982, 1987], but on a very large scale. Such scale of cargo flow from a planetary surface to circumplanetary orbits will be required & necessary for Solar-powered Planetesimals Redirection for Terraforming operations and far future Interstellar Colonization tasks.

The growing number of the PHBS lenses orbital formation infrastructure, after completing all the Planetesimals Redirection for Terraforming operations and configuring the Planetary System to maximal carrying capacity and Expansive Potential, can be used for providing concentrated beamed solar power for far future Interstellar Colonization missions, and other less important tasks.

The farther from the Sun - the more numerous and the more rich in volatiles Planetesimals are [Delsanti & Jewitt, 2006], the lower their orbital velocity and relatively cheaper in energy for delivery, but taking longer time. So, ways to increase pointing & tracking precision over interplanetary distances by 1-3+ orders of magnitude from achievable with modern technologies, can make significantly more volatile rich TNO Planetesimals available for delivery to import the limiting chemical elements for Terraforming Mars, Venus, & other Biosphere Substrates, maybe even assembling Biosphere Substrates size Planets from way too small objects like dwarf planet Ceres & all the Main Asteroid Belt.

Earth surface area is $S_{pE} = 4*\pi*r^2 = 4*\pi*(6371 \text{ km})^2$ = **510 064 472 km²**, Mars surface area is $S_{pM} = 4*\pi*(3389.5 \text{ km})^2$ = **144 371 390 km²**. Minimal requirements to Terraform Mars is to make a breathable atmosphere for humans which is comparable to Earth atmospheric pressure, and sufficient hydrosphere that after vapor saturating the atmosphere there is enough hydrosphere left to irrigate crops and provide drinking water for humans and other animals on the entire planetary surface. Delivering 1 m of water precipitation on the entire surface of Mars can be near minimum requirement for Terraforming meaning full-scale habitability, but orders of magnitude lower than desired optimal habitability of Mars for humans. If we manage to make Earth size planet from Mars - the total population Mars can feed will increase at least $S_{pE}/S_{pM}$ = **510 064 472 km²/144 371 390 km² = *3.533** times, which is very significant! Mars size Planet is near the minimal size of a Biosphere Substrate [Morozov et al, 2018a], celestial body suitable for Terraforming & inhabitation by our posterity in generations. Making Mars size Planet from Ceres by Solar-powered Planetesimals Redirection operations is the minimal requirement for Terraforming Ceres, yet if we manage to make Earth size Planet from Ceres it will increase total population in can feed about ***3.533** times compared to a Mars sized planet. Just think about the value of having 1 trillion instead of 300 billions workers on a Planet during several billions years of the remaining Planetary System lifecycle, what they can produce - and compare the value of increasing environmental carrying capacity to feed **extra 700 billions people** during **1-5 billions years**, what those extra population can produce over that time! And then compare the value of those extra 700 billions people not possible without doing those extra Terraforming work, working & reproducing over 5 billion years (Earth size Planets made from Mars & Ceres are going to die much later than Venus & Earth from Sun aging stellar lifecycle) - human*hours of work they can produce is orders of magnitude bigger than human*hours amount of work for extra Solar-powered Planetesimals Redirection operations for Terraforming efforts that makes existence and maintenance of those extra population possible. Thus, the more population thus workforce we have and the more energy we can use - the higher Terraforming quality, the higher increase in Planetary Carrying capacity thus population thus workforce over the remaining Planetary System lifecycle we can provide.

Also, for a geologically dead dwarf planet like Ceres, delivering the amount of Planetesimals that makes it into Mars size and even Earth Size Planet, along the way collecting all the Main Asteroid Belt on itself - is going to melt Ceres, thus allows to create natural geodynamo induced planetary scale magnetic field if a satellite of significant mass to create tidal forces like Earth's moon is assembled from lowest value bulk leftover Planetesimal material, sufficiently strong and long-lasting to retain breathable atmosphere during most of the Planetary System lifecycle, and facilitate reproduction of humans, animals & plants, as planetary magnetic field might be significant for many organisms reproduction bioregulation [Binghi, 2002]. The same way we can heat Mars including interiors by Planetesimals impacts as by-product of the limiting chemical elements delivery, and additionally increase Deimos & Phobos in size to strengthen the Martian magnetic field by passive methods that don't require a big fraction of most precious Equatorial latitudes for solar panels powering solenoid inductors to maintain active planetary magnetic field. The advantage of passive planetary magnetic field strengthening methods by Solar-powered Planetesimals Redirection operations for Terraforming is that passive methods like heating the planet including interiors, creating inhomogeneity and assembling moons from bulk least valuable leftover planetesimal material - don't require huge surface for Solar Panels to constantly power solenoids like active planetary magnetic field generation methods require, and don't require maintenance during all or big fraction of the Planetary System lifecycle. Some scientists [Fang & Margot, 2013] suggest that Planetary Systems are dynamically packed meaning inserting more planets might cause collisions of ejections, which means we might be able only to create 4 habitable Earth size Planets in Solar System - Venus, Earth, Mars increased up to Earth size, and Ceres increased up to Mars or maybe even Earth size by Solar-powered Planetesimals Redirection for Terraforming operations. From all vast empty space - only up to 4 habitable planets that we can Terraform & inhabit, thus we must make them all maximally habitable using Solar-powered Planetesimals Redirection for Terraforming.

With large scale or Solar-powered Planetesimals Redirection for Terraforming operations while thousands PTSSs deliver thousands of Planetesimals simultaneously, powered by thousands PHBSs - it can be possible to connect all the PHBSs on their nearly circular circumstellar orbit with ultra-lightweight carbon nanotube cables in order to gain extra orbital stability, and minimize accumulation of trajectory position & attitude error over long time by the cables averaging orbital perturbations among thousands of PHBSs due to Mercury passing by and solar power beam pressure they eject constantly rotating which is necessary to track PTSSs photoreceivers. Mercury orbit length is about $5.79*10^8$ km, PHBSs can have orbit length about **$3*10^8$ km**, with thousands of PHBSs on the same orbit average distance between each can be **$10^5$ km** and with dozens thousands **$10^4$ km** between each PHBS, which can allow to stretch 2 cables one attached to top one attached to bottom of each PHBS, 2 cables spanning forward 2 cables spanning backward the orbit, each pair attached to top & bottom of each nearest PHBS. If orbital velocity of each PHBS is slightly bigger than needed to compensate for solar radiation pressure - cable can be slightly stretched, thus tension returning each PHBS closer to its circular circumstellar orbit position when experiencing any perturbation. Thus, orbital stability & precision can be also increased with scales. Also, each PHBS can beam solar power in two opposite directions PTSSs simultaneously with two 2-axis gimballed solar-pumped lasers on optic fiber collimators, in order to compensate for light pressure tilting it, thus decreasing the load on its ADCS (Attitude Determination & Control System) [Brown, 2002; Wertz et al, 2011]. Another possible solution to decrease accumulating attitude error over a long time is to switch PHBSs to different PTSSs photoreceivers more often, which can decrease the beaming angles range for each PHBS, thus also decrease power & size requirements for ADCS of each PHBS.

One worst case scenario very inefficient idea to make it technically possible to beam solar power for PTSSs in Oort Cloud, is using an intermediate PHBS, so that PHBS tracks another PHBS which tracks PTSS with the power beam, that will relax pointing & tracking angular precision requirements more than an order of magnitude, because a PHBS diameter is more than an order of magnitude bigger that a corresponding PTSS photovoltaic array diameter, thus this can allow to obtain Planetesimals from inner Oort cloud with presently available technologies. But this is a very inefficient way, because it will require launching many extra PHBSs to beyond Kuiper Belt, following PTSS with distance many times less than the distance between the two PHBSs. This might be a worst case scenario option if nothing else possible, but increase in mechanical parts principal manufacturing precision more than an order of magnitude is preferred as the much more efficient solution if possible.

It is clear that Terraforming any other Biosphere Substrate is bigger work than everything humankind has done so far during our entire history, thus will require all humankind with orders of magnitude more population thus workforce and means of production to do Terraforming in reasonable time to work purposefully together over many generations, during a few hundreds years at least [Hossain et al, 2015; Brandon, 2018] - but imagine the value of a new Earth and its population!

As material is abundant on outer orbits of Planetary Systems [Deeg et al, 2018; Marino, 2018; Pavlenko et al, 2021; Strom, 2024; Morbidelli, 2006; Jewitt et al, 1998, 2001, 2004; Delsani et al, 2006], the main principal limiting factor determining technically possible quality of Terraforming, is the amount of energy available & distance we can precisely enough beam it to. Power is almost unlimited if we manage to use beamed concentrated starlight as the power source for Ion Thrusters & other systems, every star irradiates orders of magnitude more power that we can use, with the same rate regardless of how much we use stellar power. Ability to scale up humankind and our posterity particularly as much as physically possible is the main motivation to develop Solar-powered Planetesimals Redirection for Terraforming technologies, even though they are technologically more complex than nuclear, but allow orders of magnitude bigger scale of Terraforming quality = Planetary Carrying Capacity increase.

## 6. Conclusions

We calculated & found that it is principally physically & technically possible to conduct Solar-powered Planetesimals Redirection for Terraforming operations by PTSSs using in-situ evaporated Planetesimal material as-is as propellant for Ion Thrusters and powered by photovoltaic panels arrays receiving concentrated precisely beamed solar power from PHBSs orbiting a bit closer than the orbit of Mercury tracking PTSSs photoreceivers during the entire missions, delivering volatile rich TNO Planetesimals from Kuiper Belt to Mars, Venus, and similar Exoplanets with presently available materials & technological processes, and from Oort Cloud with increasing principally available mechanical parts manufacturing precision by 1-3+ orders of magnitude.

We can conclude that Solar-powered Planetesimals Redirection for Terraforming follows the economy of scales, becoming technologically easier & more efficient with growing scales.

The main purpose of the article is to scientifically prove the concept principal physical and technical feasibility and initiate the scientific discussion on how to further develop scientific and technical basis that can efficiently and productively enable this class of Solar-powered Planetesimals Redirection for Terraforming operations in the future, involve more people in the development. The topic is too big to fit into one paper: this paper was intended only to address the concept principal physical and technical feasibility issues and make preliminary results sooner available to global general public as we already worked during several years on it - here we omitted many topics that might increase the total cost a few % but not orders of magnitude thus are not principal physical and technical feasibility issues, we just did rough orders of magnitude estimations enough to prove principal feasibility. This is the first paper in the planned series of several peer-reviewed journal publications and PhD Thesis on Solar-powered Planetesimals Redirection for Terraforming, 3rd paper is planned about detailed design of PTSS and 4th about detailed design of PHBS both including more detailed materials science calculations, 5th paper planned about detailed orbital mechanics mission planning, 6th about controls telecommunications and a number of less important aspects.

As we might lack some knowledge in many areas overlapping the proposed class of technological schemes and its principal physical limitations, we invite any specialists with relevant knowledge to verify our assumptions, to check for more principal physical & technical limitations if any, suggesting ways & approaches to overcome them, and to develop the Solar-powered Planetesimals Redirection for Terraforming topic further in more details in growing scales both in discussion with us and in independent publications on any aspects of the mentioned class of operations development.

Near-term small-scale experimental verification of the described concept technical feasibility most importantly must maintain the relative proportion between orbital beaming distances, planetesimals and the spacecrafts sizes and masses.

**Acknowledgements**

We are grateful to Dr. Robert Zubrin for the useful long in-person discussion on the Terraforming topic at the 42nd COSPAR Scientific Assembly in Pasadena, California, USA.

We are grateful to the Institute of Biophysics SB RAS, Laboratory of Controlled Biosynthesis of Phototrophic Organisms, for allowing the main author to be part time employed on relevant research job during many years and have a lot of free time to study & research Terraforming & related topics without which this work won't be possible, and for useful discussions about biosphere dynamics & Terraforming, especially Prof. Dr. Sergey Trifonov.

We thank Artem Morozov for drawing timely Figure 5 main infographics of the Solar-powered Planetesimals Redirection for Terraforming concept overview.

We are grateful to Prof. Dr. Kyle Vaught & Prof. Dr. James Lewis for useful discussions on the topic & teaching relevant classes for the main author.